\documentclass[showpacs,prl,onecolumn,aps,superscriptaddress,preprintnumbers,nofootinbib]{revtex4}
\usepackage[T1]{fontenc}
\usepackage[latin9]{inputenc}
\usepackage{amsmath,amssymb}
\usepackage{epsfig}
\usepackage{bm} 
\usepackage{graphicx}
\usepackage{mathrsfs}
\usepackage{amsmath}
\usepackage{amsfonts}
\usepackage{epstopdf}
\usepackage{color}
\usepackage{pifont}
\usepackage{relsize}
\usepackage{mathtools}
\def\slashchar#1{\setbox0=\hbox{$#1$}     		
   \dimen0=\wd0                                 	
   \setbox1=\hbox{/} \dimen1=\wd1               	
   \ifdim\dimen0>\dimen1                        	
      \rlap{\hbox to \dimen0{\hfil/\hfil}}      	
      #1                                        	
   \else                                        	
      \rlap{\hbox to \dimen1{\hfil$#1$\hfil}}   	
      /                                         	
   \fi}
\newcommand{\gaa}{\overline{\gamma}}

\renewcommand{\vec}{\boldsymbol}
\newcommand{\beq}{\begin{equation}}
\newcommand{\eeq}{\end{equation}}
\newcommand{\bea}{\begin{eqnarray}}
\newcommand{\eea}{\end{eqnarray}}
\newcommand{\baa}{\begin{array}}
\newcommand{\eaa}{\end{array}}

\def\eq#1{{Eq.~(\ref{#1})}}

\def\fig#1{{Fig.~\ref{#1}}}

\newcommand{\bas}{\bar{\alpha}_S}

\newcommand{\De}{\Delta}

\newcommand{\Lb}{\left(}
\newcommand{\Rb}{\right)}

\renewcommand{\vec}[1]{\boldsymbol{#1}}
\newcommand{\dif}{\mathrm{d}}

\begin{document}
\title{CGC/saturation approach: Impact-parameter dependent model of perturbative QCD and combined HERA data}
\author{Michael Sanhueza}
\email{michael.roa@upla.cl}
\affiliation{Facultad de Ingenier\'ia, Laboratorio DataScience, Universidad de Playa Ancha, Leopoldo Carvallo 270, Valpara\'iso, Chile}
\affiliation{Centro de Estudios Avanzados, Universidad de Playa Ancha, Traslavi\~{n}a 450, Vi\~{n}a del Mar, Chile}

\author{Jos\'e Garrido}
\email{jose.garridom@sansano.usm.cl}
\affiliation{Departamento de F\'isica, Universidad T\'ecnica Federico Santa Mar\'ia,  Avda. Espa\~na 1680, Casilla 110-V, Valpara\'iso, Chile}

\author{Miguel Guevara}
\email{miguel.guevara@upla.cl}
\affiliation{Facultad de Ingenier\'ia, Laboratorio DataScience, Universidad de Playa Ancha, Leopoldo Carvallo 270, Valpara\'iso, Chile}


\keywords{CGC/saturation approach, impact parameter dependence
 of the scattering amplitude, solution to non-linear equation, deep inelastic
 structure function, diffraction at high energies}
\pacs{12.38.Cy, 12.38g,24.85.+p,25.30.Hm}

\begin{abstract}
In this paper we confront the CGC/saturation approach of Ref.~\cite{CLMS} with the experimental combined HERA data and obtain its parameters. The model includes two features that are in accordance with our theoretical knowledge of deep inelastic scattering. These consist of: \textit{i}) the use of analytical solution for the non-linear Balitsky-Kovchegov (BK) evolution equation and \textit{ii}) the exponential behavior of the saturation momentum on the impact parameter $b$-dependence, characterized by $Q_s$ $\propto\exp\Lb -m b \Rb$ which reproduces the correct behavior of the scattering amplitude at large $b$ in accord with Froissart theorem. The model results are then compared to data at small-$x$ for the structure-function of the proton $F_{2}$, the longitudinal structure function $F_{L}$, the charm structure function $F_2^{c\bar{c}}$, the exclusive vector meson ($J/\psi,\phi,\rho$) production and Deeply Virtual Compton Scattering (DVCS). We obtain a good agreement for the processes in a wide kinematic range of $Q^2$ at small-$x$. Our results provide a strong guide for finding an approach, based on Color Glass Condensate/saturation effective theory for high energy QCD, to make reliable predictions from first principles and for forthcoming experiments like the Electron-Ion Collider and the LHeC. 
\end{abstract}

\maketitle

\vspace{-0.5cm}



\section{ Introduction}
The goal of this paper is to confront the Color Glass Condensate (CGC)/saturation approach of Ref.~\cite{CLMS} with the experimental combined HERA data and obtain its parameters via fit. To contrast with experimental data,  it was selected deep inelastic processes using the CGC/saturation equations from Ref.\cite{JIMWLK} in a simple version of the non-linear Balitsky-Kovchegov (BK) equation \cite{BK}. 
The authors of Ref.~\cite{CLMS} proposed a way to take into account corrections, which coincides with other attempts on the market, as far as the linear dynamics is concerned, but introduce the non-linear evolution which guarantees the correct high energy asymptotic behavior of the scattering amplitude. They include the re-summation procedure suggested in  Refs.~\cite{SALAM,SALAM1,SALAM2}, to fix the BFKL kernel. Specifically, the rapidity variable was introduced in the same way as in Ref.~\cite{DIMST}. Nevertheless, a different way to account for these corrections was suggested which leads to additional changes in the evolution equation. The advantage of the BFKL kernel ~\cite{BFKL,LIP},  is that it leads to the scattering amplitude satisfying high-energy limits, which follows from the approach of Ref.~\cite{LETU} (see Refs.~\cite{CLMP,XCWZ}) concerning the BK ~\cite{BK} evolution equation ~\cite{NLOBK0,NLOBK01,NLOBK1,NLOBK2,JIMWLKNLO1,JIMWLKNLO2,JIMWLKNLO3}.

One of the major challenges in high-energy scattering theory is finding a reliable description of the non-linear evolution of scattering processes. The leading-order CGC approach, while providing a framework, predicts an energy dependence for the scattering amplitude and saturation momentum that leads to growth at high energies. This growth behavior contradicts existing experimental data. To address this discrepancy, theoretical efforts have focused on incorporating corrections beyond the leading order. In Ref.~\cite{CLMS}, a significant step forward is made by proposing an analytical solution to the leading-order (LO) non-linear Balitsky-Kovchegov (BK) equation. By incorporating key features relevant to high-energy scattering processes, this solution has the potential to provide a more accurate description of the non-linear evolution, leading to better agreement with HERA data.

The model which is based on CGC/saturation effective theory for high energy QCD (see Ref.~\cite{KOLEB} for a review), includes the impact parameter dependence of the scattering amplitude. Within this framework, the scattering amplitude falls down at large impact parameters $b$ as a power of $b$. Such a power-like decrease leads to the violation of the Froissart theorem ~\cite{FROI}. Therefore, we have no choice but to build a model because CGC/saturation equations cannot reproduce the correct behavior of the scattering amplitude at large impact parameters \cite{KW,FIIM}. 
In accordance with the geometric scaling behavior of the scattering amplitude~\cite{GS, SGBK} and guided by the semi-classical solution to the CGC/saturation equations~\cite{BKL}, this approach incorporates the non-perturbative impact parameter behavior into the saturation momentum. In most cases, these models include the non-perturbative $b$-behavior of the scattering amplitude and has been widely applied in what is commonly referred to as saturation models~\cite{SATMOD0,SATMOD01,SATMOD1,SATMOD2,IIM,SATMOD3,SATMOD4,SATMOD5,SATMOD6,SATMOD7,SATMOD8,SATMOD9,SATMOD10,SATMOD11,SATMOD12,SATMOD13,SATMOD14,SATMOD15,SATMOD16,SATMOD17,CLP,CLMP,CLS}. 
In the choice of the nonperturbative behavior of the saturation scale, instead to use the common approach  proportional to $\exp \Lb -b^2/B\Rb$ which disagrees the theoretical knowledge of QCD at high energies ~\cite{FROI,LETU},
we will parametrize as follows~\cite{CLP,CLMP,CLS}: 
\beq \label{QSB}
 Q^2_s\Lb b , Y\Rb \,\propto\, \Lb S\Lb b, m \Rb\Rb^{\frac{1}{\bar \gamma}}
\eeq
where $S\Lb b \Rb $ is the Fourier  image of $ S\Lb Q_T\Rb = 1/\Lb 1 +    \frac{Q^2_T}{m^2}\Rb^2$ and the value of $\bar \gamma$ we will discuss in section III.  In the vicinity of the saturation scale, this $b$ dependency leads to a significant $b$-dependence of the scattering amplitude, which is proportional to $\exp\left( - m b\right)$ for $b \gg 1/m$, consistent with the Froissart theorem~\cite{FROI}. An important aspect, the model's $b$-dependence is congruent with perturbative QCD calculations for large values of momentum transferred ($Q_T$), leading to a power-like behavior in the scattering amplitude at high $Q_T$~\cite{LEBR}.

Based on these observations and remarks, this paper undertakes an examination of how the CGC/saturation dipole model reproduces the combined HERA data and conducts a detailed analysis thereof.  Furthermore, the combined data for inclusive deep inelastic scattering (DIS) have been measured with high accuracy by H1 and ZEUS collaboration \cite{HERA1,HERA2} featuring extremely small error bars. This characteristic underscores the substantial challenge faced by any theoretical approach in accurately describing this data. Fitting the model to this data allows us to extract all phenomenological parameters that are introduced in the model. After determining the values of all phenomenological free parameters we calculated and compared the theoretical results with experimental data at small-$x$ for several observables, including the proton structure function $F_{2}$, the longitudinal structure function $F_{L}$, the charm structure function $F_2^{c\bar{c}}$, as well as exclusive vector meson production ($J/\psi,\phi,\rho$)  and Deeply Virtual Compton Scattering (DVCS). 

Our results provide a strong guide in the continuous process for finding an approach, based on Color Glass Condensate/saturation effective theory for high energy QCD, to make reliable predictions from first principles, as well as for forthcoming experiments like the Electron-Ion Collider~\cite{EIC} and the LHeC~\cite{LHEC}. Another important application is to use the dipole amplitude for the production of dijets in p-p and p-Pb collisions~\cite{KUSA} or the structure of the soft Pomeron in CGC~\cite{CLS2}.
This paper is organized as follows. Section II introduces the framework for computing the total DIS cross-section, structure functions, and exclusive diffractive processes within the color dipole approach. Section III details the CGC/saturation dipole model employed for describing the experimental data. We present the key aspects of the formulation and introduce the phenomenological parameters that need to be calculated within the non-perturbative QCD framework. Section IV discusses the numerical results obtained from our analysis.  Finally, the conclusion summarizes our findings.

\section{Description of inclusive and exclusive diffractive processes}
\subsection{Inclusive processes: Total DIS cross-section and proton structure functions}
The observables in deep inelastic scattering (DIS) can be expressed through the following scattering amplitudes (see Ref.~\cite{KOLEB} and references therein):

\beq\label{FORMULA}
N_{L,T}\Lb Q, Y; b\Rb \,\,=\,\,\int \frac{d^2 r}{4\,\pi} \int^1_0 d z \,|\Psi^{\gamma^*}_{L,T}\Lb Q, r, z\Rb|^2 \,N\Lb r, Y; b\Rb
\eeq
where $Y \,=\,\ln\Lb 1/x_{Bj}\Rb$ and $x_{Bj}$ is the Bjorken $x$.  $z$ is the fraction of the light cone momentum of the virtual photon carried by quark. $Q$ is the photon virtuality and $L,T$ denote the longitudinal and transverse polarizations of the virtual photon. 
In the dipole picture of DIS at high energies, the interactions are characterized by the features shown in \eq{FORMULA}. The process occurs as follows: in the first stage the virtual photon decay into quark-antiquark pair, described by $|\Psi^{\gamma^*}_{L,T}\Lb Q, r, z\Rb|^2$. During the second stage, the color dipole proceeds to interact with the target (proton) represented by $N(r,Y;b)$, which is the imaginary part of the forward $q\bar{q}$ dipole-proton scattering amplitude with transverse dipole size $r$ and impact parameter $b$.

The wave function for $|\Psi^{\gamma^*}_{L,T}(Q,r,z)|^2\equiv(\Psi^*\Psi)^{\gamma^*}_{L,T}$ is well known (see Ref.~\cite{KOLEB} and references therein).
\begin{align}
  (\Psi^*\Psi)_{T}^{\gamma^*} &=
   \frac{2N_c}{\pi}\alpha_{\mathrm{e.m.}}\sum_f e_f^2\left\{\left[z^2+(1-z)^2\right]\epsilon^2_f K_1^2(\epsilon_f r) + m_f^2 K_0^2(\epsilon_f r)\right\},\label{WFDIST}   
  \\
  (\Psi^*\Psi)_{L}^{\gamma^*}&
  = \frac{8N_c}{\pi}\alpha_{\mathrm{e.m.}} \sum_f e_f^2 Q^2 z^2(1-z)^2 K_0^2(\epsilon_f r),
\label{WFDISL}
\end{align}
with
\beq\label{EPSILON}
\epsilon^2_f\,\,=\,\,m^2_f\,\,+\,\,z (1 - z) Q^2
\eeq
where $\alpha_{\mathrm{e.m.}}$ is the electromagnetic fine structure constant, $N_c$ denotes the number of colors and $e_f$ is the electric charge of a quark with flavor $f$ and mass $m_f$. In $c\bar{c}$ scattering, the mass of the charm quark (about $m_c=1.4 \,\rm{GeV}$) is not small and we took this into account by replacing $x$ as $x_c=x_{Bj}(1+4m^2_c/Q^2)$. 
 Using \eq{FORMULA}, \eq{WFDIST} and \eq{WFDISL}  we can write the main observables in DIS as follows:
\begin{subequations}
 \bea 
 \sigma_{T,L}^{\gamma^*p}(Q^2,x) &\,\, =\,\,& 2 \int d^2 b \,\,N_{T,L}\Lb Q,Y;b\Rb; \label{SIGMA}\\
 F_2\Lb Q^2, x\Rb &=& \frac{Q^2}{4\,\pi^2\,\alpha_{\rm e.m.}} \big[\sigma_T^{\gamma^*p}(Q^2,x)\,\,+\,\,\sigma_L^{\gamma^*p}(Q^2,x)\big];\label{F2}\\
  F^{c\bar{c}}_2\Lb Q^2, x\Rb &=& \frac{Q^2}{4\,\pi^2\,\alpha_{\rm e.m.}} \Big[\sigma^{c\bar{c},\gamma^*p}_T(Q^2,x)\,\,+\,\,\sigma^{c\bar{c},\gamma^*p}_L(Q^2,x)\Big];\label{FCC}\\
  F_L\Lb Q^2, x\Rb &=& \frac{Q^2}{4\,\pi^2\,\alpha_{\rm e.m.}} \,\sigma_L^{\gamma^*p}(Q^2,x);\label{L} 
 \eea
\end{subequations} 
The reduced cross-section $\sigma_r$ is expressed in terms of the inclusive proton structure functions $F_2$ and $F_L$ as
\beq \label{SIGMAR}
\sigma_r(Q^2,x,y)=F_2(Q^2,x)-\frac{y^2}{1+(1-y)^2}F_L(Q^2,x)
\eeq
where $y=Q^2/(sx)$ denotes the inelasticity variable and $\sqrt{s}$ indicates the center of mass energy in ep collisions.

\subsection{Exclusive diffractive processes: Deeply virtual Compton scattering and exclusive vector meson production}

In the dipole approach, the scattering amplitude for the exclusive diffractive processes $\gamma^{*} + p \rightarrow E + p$, with the final state a real photon $E = \gamma$ in DVCS or vector meson $E = J/\psi,\phi, \rho$ can be written in terms of a convolution of the dipole amplitude $N$ and the overlap wave functions of the photon and the exclusive final-state particle. The main formula for exclusive diffractive processes takes the form~\cite{SATMOD6}:

\beq\label{EXCLUSIVE}
{\cal{A}}^{\gamma^{*}p\, \rightarrow \,E p}_{T,L}(x,Q,\De)=2i \int d^{2}\vec{r} \int_{0}^{1} \frac{dz}{4\pi} \int d^{2}\vec{b}  \ (\Psi_{E}^{*}\Psi)_{T,L}
\ \mathrm{e}^{-i[ \vec{b}-(\frac{1}{2}-z)\vec{r}]\cdot\vec{\De}}\,N\Lb r, Y; b\Rb
\eeq

where $|\vec{\De}|^{2} =-t$, and $t$ represents the squared momentum transfer. The phase factor $\exp(i(\frac{1}{2}-z)\vec{r}\cdot\vec{\De})$ is due to the non-forward wave-functions contribution~\cite{BGBP}, with the correct phase factor of $\frac{1}{2}$ found in~\cite{HXY}.

In terms of the scattering amplitude of \eq{EXCLUSIVE}, the differential cross-section for exclusive diffractive processes may be written as~\cite{SATMOD3,SATMOD6}
\beq\label{DSIGMA}
\frac{d\sigma_{T,L}^{\gamma^{*}p \, \rightarrow \,E p}}{dt}\,=\,\frac{1}{16\pi}\left|{\cal{A}}^{\gamma^{*}p\, \rightarrow \,E p}_{T,L}\right|^{2} \Lb 1+\beta^2\Rb\,
\eeq
where
\beq\label{SIGMAEXC}
\sigma^{\gamma^* p\rightarrow E p}_{T,L}  = \int d t   \frac{\dif\sigma^{\gamma^* p\rightarrow E p}_{T,L}}{\dif t}
\eeq
and 
\beq\label{B_D}
B_D=\lim_{t\rightarrow 0}\,\frac{d}{dt}\ln\,\Lb\frac{d\sigma_{T,L}^{\gamma*p\rightarrow E p}}{dt}\Rb
\eeq
are the two main observables that we are going to use to calculate our theoretical estimates.
In order to account for the missing real part of \eq{EXCLUSIVE}, a multiplicative factor $\Lb1+\beta^2\Rb$ must be added, where $\beta$ is the ratio of the real to imaginary parts of the scattering amplitude which shows Regge-type behavior at high energies~\cite{GLM,NNPZ,MRT}
\beq\label{BETAFACTOR}
\beta=\tan(\pi\delta/2)\;,\;\;\mathrm{with}\;\;\delta\equiv\frac{\partial\ln({\cal{A}}^{\gamma^{*}p\, \rightarrow \,E p}_{T,L})}{\partial\ln(1/x)}
\eeq
It is worth mentioning that for exclusive diffractive processes, it becomes imperative to consider the skewedness effect. This arises since the gluons attached to the $q\bar{q}$ pair can carry distinct light-cone fractions $x,x'$ of the proton. At the next-to-leading order (NLO) level, particularly under the condition where $x'\ll x \ll1$, the skewedness effect~\cite{SGBMR} can be accounted by simply  multiplying the gluon distribution $xg(x, \mu^2)$ by a factor $R_g$ defined via
\beq\label{RGFACTOR}
R_g=\frac{2^{2\gamma+3}}{\sqrt{\pi}}\frac{\Gamma(\gamma+5/2)}{\Gamma(\gamma+4)}\;,\;\;\mathrm{with}\;\;\gamma\equiv\frac{\partial\ln[xg(x,\mu^2)]}{\partial\ln(1/x)}
\eeq
For deeply virtual Compton scattering (DVCS) the overlap wave function only has a transversal contribution. Similar to \eq{WFDIST} where a sum over quark flavors should be done, the equation read as

\beq\label{WFDVCS}
(\Psi_{\gamma}^{*}\Psi)^{(DVCS)}_{T}=\frac{2 N_c}{\pi}\alpha_{\mathrm{em}}\sum_f e_f^2\{[z^2+(1-z)^2]\epsilon_fK_1(\epsilon_fr)m_fK_1(m_fr)+m_f^2K_0(\epsilon_fr)K_0(m_fr)\}
\eeq

For vector meson diffractive production, we have wave functions exclusively for mesons composed of heavy quarks. The $J/\psi$ is the most commonly cited example. Nevertheless, it is important to take into account that the charm quark's mass is not very large, and corrections can be crucial. For all other mesons, confinement corrections are substantial, and the wave function, motivated by heavy quark mesons, can only be considered as pure phenomenological assumptions. The overlap integrals are represented according to the prescription from Ref. \cite{SATMOD3}
\begin{align}
(\Psi_{V}^{*}\Psi)_{T}&=\hat{e}_fe\frac{N_c}{\pi z(1-z)}\{m_f^2K_0(\epsilon_fr)\phi_T(r,z)-[z^2+(1-z)^2]\epsilon_fK_1(\epsilon_fr)\partial_r\phi_T(r,z)\}\label{WFDVMP1}
\\
(\Psi_{V}^{*}\Psi)_{L}&=\hat{e}_fe\frac{N_c}{\pi}\,2\,Qz(1-z)K_0(\epsilon_fr)\Big[M_V\phi_L(r,z)+\frac{m_f^2-\nabla_r^2}{M_Vz(1-z)}\phi_L(r,z)\Big] \label{WFDVMP2}
\\
\phi_{T,L}(r,z)&=\mathcal{N}_{T,L}z(1-z)\exp\bigg(-\frac{m_f^2\mathcal{R}^2}{8z(1-z)}-\frac{2z(1-z)r^2}{\mathcal{R}^2}+\frac{m_f^2\mathcal{R}^2}{2}\bigg)
\label{WFDVMP3}
\end{align}

where $\nabla^2_r\equiv(1/r)\partial_r+\partial_r^2$, $M_V$ is the meson mass and the effective charge $\hat{e}_f$ takes the values of $2/3$, $1/3$, or $1/\sqrt{2}$ for
$J/\psi$, $\phi$, or $\rho$ mesons respectively. The parameters $\mathcal{N}_{T,L}$, $\mathcal{R}$ and $m_{f}$ are from Table 2 in Ref.~\cite{SATMOD3}.

\section{CGC/saturation dipole model}
\subsection{$q\bar{q}$ dipole-proton scattering amplitude}
To compute the total cross-section, the proton structure functions in DIS, exclusive diffractive vector meson production, and Deeply Virtual Compton Scattering,  we need to use the $q\bar{q}$ dipole-proton forward scattering amplitude.  In Ref.~\cite{CLMS} the analytical solution for the nonlinear BK equation was found at the NLO BFKL kernel in the saturation domain.  The authors propose a method to incorporate these corrections by introducing non-linear evolution, which ensures the correct high-energy asymptotic behavior of the scattering amplitude. They incorporate the re-summation procedure proposed in  Refs.~\cite{SALAM,SALAM1,SALAM2}, to address the BFKL kernel at higher orders. Specifically, the rapidity variable was introduced in a similar manner that in Ref.~\cite{DIMST}. 
Nevertheless,  a different way to account the non-linear corrections was suggested: the anomalous dimension was derived in the region of large $\tau=r^2Q_s^2$ in the limit when $\gamma \to 0$ by using the kernel in $\gamma$-representation for the leading twist in place of the full BFKL kernel (see equation 90 from Ref.~\cite{CLMS}), which corresponds to the sum of two types of logarithms~\cite{LETU}. The advantage of the BFKL kernel ~\cite{BFKL,LIP},  is that it leads to the scattering amplitude satisfying high-energy limits, which follows from the approach of Ref.~\cite{LETU} (see Refs.~\cite{CLMP,XCWZ}) concerning to the NLO-BK~\cite{BK} evolution equation~\cite{NLOBK0,NLOBK01,NLOBK1,NLOBK2,JIMWLKNLO1,JIMWLKNLO2,JIMWLKNLO3}.

In the CGC/saturation dipole model, the color $q\bar{q}$ dipole-proton scattering amplitude is given by ~\cite{CLMS,CLS}: 

\bea \label{NZ}
N\Lb z\Rb\,\,=\,\, \left\{\begin{array}{l}~~~~~~~~~~~~~~~~~~~~N_0\,e^{z\bar{\gamma}}\ \ \ \ \ \ \ \ \ \ \ \ \ \ \ ~~~~~~~~~~~ \,\mbox{for}\ \  \ \ \tau\,\leq\,1;\\\ \\
\,\,\,a\,\Big( 1\,\,-\,\,e^{- \Omega\Lb z \Rb}\Big) \,\,+\,\,\Lb 1\,-\,a\Rb\frac{\Omega\Lb z \Rb}{1\,\,+\,\,\Omega\Lb z \Rb}\,\,\,\,\,~~~~\mbox{for} \ \  \ \ \tau\,>\,1;\\  \end{array}
\right.
\eea
The parameter $a = 0.65$ describes the exact solution of the nonlinear BK equation within an accuracy of less than 2.5\%~\cite{LEPP, CLS}, while the function $\Omega(z)$ takes the following form according to references~\cite{CLMS,CLS}:

\beq \label{OMEGA}
\Omega(z)\,\,=\,\,\Omega _0	\Bigg\{ \cosh \left(\sqrt{\sigma } z\right)\,\,+\,\,\frac{\bar{\gamma}}{\sqrt{\sigma}}\, \sinh \left(\sqrt{\sigma} z\right)\Bigg\},~~~~~~\mathrm{with}\;\;\sigma=\frac{\bas}{\lambda(1+\bas)};
\eeq

The parameter $\Omega _0$ is numerically equivalent to $N_0$, ensuring the correct behavior of the solution for \eq{NZ}\footnote{This equivalence is maintained because $\Omega_0$ is a small parameter, making $N_0$ approximately equal to $\Omega_0$. This relationship is essential for ensuring consistency in the model's predictions.}. $N_0$ represents the value of the scattering amplitude at $\tau=1$ and $\bas = N_c\,\alpha_S/\pi$ where $N_c$ denotes the number of colours. While theoretically, the value of $N_0$ can be computed using the linear evolution equation with appropriate initial conditions, it inherently relies on the phenomenological parameters of said initial condition. Consequently, we treat $N_0$ as a parameter to be determined through fitting. 

The variable $z$ is defined as
\beq \label{xiformula}
 z=\ln\,(r^2Q^2_s(Y,b))
\eeq 
The use of this variable indicates the main idea of the approach in the region for $\tau=r^2Q_s^2 > 1$. The goal is to establish a correspondence between the solution of the non-linear equation and the solution of the linear equation in the kinematic region described by $N=N_0\,e^{z\bar{\gamma}}$.

The critical anomalous dimension and the energy behavior of the saturation scale were determined in  Ref.~\cite{CLMS}  which has the following form:
\beq \label{gammaold}
\bar{\gamma}=\bar{\gamma}_\eta(1+\lambda_\eta),~~~~~~~~~\mathrm{with}\;\;\bar{\gamma}_\eta=\sqrt{2+\bas}-1;
\eeq 
and
\beq \label{lambda}
    \lambda=\dfrac{\lambda_\eta}{1+\lambda_\eta},~~~~\mathrm{with}\;\;\lambda_\eta=\dfrac{1}{2}\frac{\bas}{\Lb3+\bas-2\sqrt{2+\bas}\Rb};\;\;\;\;
\eeq
where $\eta=Y-\xi$ is a new energy variable, with $Y=\ln(1/x)$ the rapidity of the dipole and $\xi=\ln(r^2Q_s^2(Y=0,\mathbf{b}))$. 

We use a different way to introduce the parameters from the experimental data: expanding the linear solution of \eq{NZ} to the region $\tau<1$, replacing $\gaa$ by the following expression:

\beq \label{gammanew}    \bar{\gamma}\rightarrow\bar{\gamma}+\frac{\ln(1/\tau)}{2\kappa\lambda Y},\;\;\mathrm{with}\;\;\kappa=\frac{\chi''(\bar{\gamma})}{\chi'(\bar{\gamma})}=\frac{\frac{d^2\omega(\bar{\gamma}_\eta)}{d\bar{\gamma}_\eta^2}}{\frac{d\omega(\bar{\gamma}_\eta)}{d\bar{\gamma}_\eta}}
\eeq
this equation was derived in Ref.~\cite{IIML} and the results from saturation models~\cite{SATMOD0,SATMOD1,SATMOD2,IIM,SATMOD3,SATMOD4,SATMOD5,SATMOD6,SATMOD7,SATMOD8, SATMOD9,SATMOD10,SATMOD11,SATMOD12,SATMOD13,SATMOD14,SATMOD15,SATMOD16,SATMOD17,CLP,CLMP,CLS} demonstrate  a strong agreement with the experimental data for $x\leq0.01$. The eigenvalue of $\omega(\gaa_{\eta})$ was taken from Ref.~\cite{CLMS}.

\subsection{Phenomenological input: Impact parameter dependence of the saturation scale}
As a first step, we have only incorporated one phenomenological parameter:  $N_0$ which determines the scattering amplitude at $r^2\,Q^2_s = 1$.  However,  we must define the initial conditions at $Y = 0$ for the linear evolution equation in the region where $\tau = r^2Q_s^2\leq 1$. This requires specifying the saturation scale at impact parameter $b = 0$ and its dependence on $b$.  We recognize that the confinement of quarks and gluons at large $b$ is crucial for the amplitude's exponential fall-off. However, both aspects are constrained to the non-perturbative realm of QCD, where our current understanding is limited, leading to the suggestion of phenomenological parameterization. We use the following expression from Refs.~\cite{CLP, CLMP} to represent the saturation momentum:
\bea
Q^{2}_s\Lb Y, b\Rb\,\,&=&\,Q^{2}_s\Lb Y=0,b=0\Rb\,\,S(b)^{1/\bar{\gamma}}\,\,e^{\lambda\,Y} ~~=\,\,Q^{2}_0\,(m\,b\, K_{1}(m\,b))^{1/\bar{\gamma}}\,e^{\lambda\,Y};\label{QS1}
\eea
This expression leads to  $Q^2_s\Lb Y = Y_0,b\Rb\,\xrightarrow{m b \,\gg\,1} \exp\Lb - m b/\bar{\gamma}\Rb$, differs from others by providing the correct large $b$ behavior of the scattering amplitude. The exponential decrease at large $b$ adhering to theoretical principles of analyticity and unitarity~\cite{FROI}. This approach stands in contrast to models from Refs.~\cite{SATMOD5,SATMOD6,SATMOD7,SATMOD8,SATMOD9,SATMOD12,SATMOD17} using $Q^2_s\Lb Y = Y_0,b\Rb\,\propto \exp\Lb- b^2/B\Rb$.  
How the scattering amplitude behaves at large impact parameter $b$ is critical for understanding how the interaction radius changes with collision energy. According to a fundamental principle (Froissart theorem~\cite{FROI}), $b$  should increase proportionally with $Y$ to ensure the interaction radius has the correct energy dependence. This difference in $b$-dependence leads to distinct relationships between the interaction radius (R) and $Y$: with an exponential decrease, R scales as $ (1/m) Y$, while a Gaussian decrease results in R scaling as  $ (1/m) \sqrt{Y}$ \footnote{Parametrized models such as IP-Sat and b-CGC, alongside hotspot models, utilize the Gaussian distribution in $b$ due to its simplicity and effective fit to experimental data. Research from Refs. \cite{MS,S,KUTOLL} have demonstrated that these Gaussian profiles can evolve with energy while adhering to the Froissart bound. However, perturbative calculations and the study of large $b$ behavior in QCD suggest that $\exp(-mb)$ might be more suitable to ensure theoretical consistency in high-energy collisions.}. This distinction results in a rapid increase of the scattering amplitude in this particular parameterization, influencing predictions at high energy. The proposed equation for $Q_s^2$ gives the amplitude close to the saturation scale, proportional to $S(b)$,  and generates the behavior $ 1/\Lb 1 + \frac{Q^2_T}{m^2}\Rb^2$, where $Q_T$ is the momentum transfer. At large $ Q_T$, the amplitude in the model is proportional to $ 1/Q^4_T$, in line with perturbative QCD calculations~\cite{LEBR}, a behavior not reproducible with a Gaussian distribution.

\begin{figure}
\centering 
   \includegraphics[width=10cm]{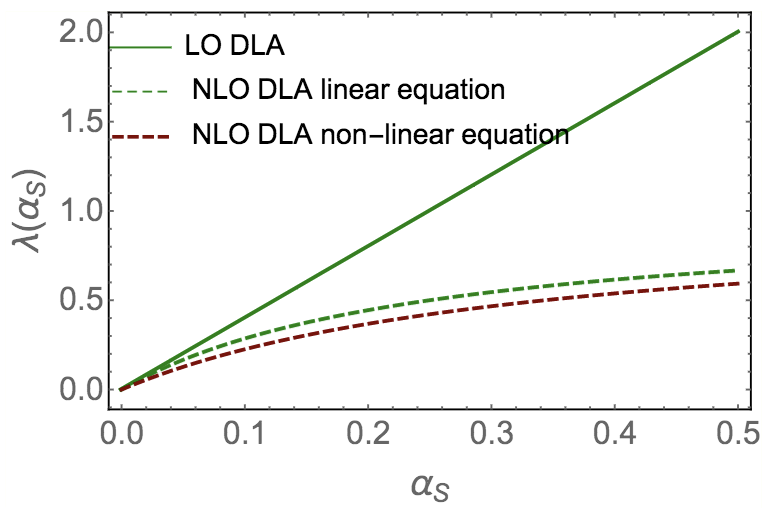}
 \caption{$\lambda$ for $Y$ dependence of the saturation scale: $Q^2_s(Y)=Q^2_s(Y=0)\exp(\lambda Y)$ versus $\bas$. The solid line
corresponds to $\lambda=4\bas$ in leading order (LO) of double log approximation (DLA), the dashed green curve describes the NLO DLA value of the linear equation and the dashed red curve indicates the NLO DLA value of non-linear equation in the considered kinematic region. The figure was taken from Ref.~\cite{CLMS}.}  
\label{LAMBDANLO}
\end{figure}

The values of $Q^{2}_0$ and $m$ will be determined through a fitting procedure. Previous fitting results have yielded $Q^{2}_0$ values ranging from $0.15$ to $0.25 \,\rm GeV^2$, and $m$ values approximately between $0.4$ and $0.85\,\rm{GeV}$.  This range encompasses  $m=0.72\, \rm{GeV}$, which serves as the scale for the electromagnetic form factor of the proton, while $m\,\approx\,0.5 \,\rm{GeV}$ corresponds to the scale for the so-called gluon mass~\cite{CLP,CLMP}.
$\lambda$ and $\gaa$ are defined using the respective expressions provided in \eq{gammaold}, \eq{lambda} and \eq{gammanew}. The value of $\lambda$ represents the energy dependence of the saturation scale, with the typical value of $\lambda=0.2$ required for accurate description in DIS data. In the model proposed in Ref.~\cite{CLMS}, it was demonstrated that for $\bas$  close to $0.1$,  the value of $\lambda$ is approximately $0.2$.  Subsequently, in Ref.~\cite{CLS}, the model was tested against experimental data with the value of $\bas$ treated as a free parameter. Remarkably, the model, even when $\bas$ was set to $0.2$, consistently reproduced the experimental data. In essence, a freely determined $\bas$ from experimental data is essential to establish $\lambda(\bas)$ at a level capable of accurately describing HERA data. It is worthwhile mentioning that the corrections to the energy dependence of the saturation momentum give considerable contributions at small values of $\bas$. Therefore, we consider $\bas$ as the fourth parameter of the model, which we expect $\bas \approx 0.10$ according to the predictions in the calculations for perturbative QCD (see  \fig{LAMBDANLO}). In summary, we include three phenomenological parameters originating from the initial conditions, and their values are obtained through the fitting procedure with experimental data.

\begin{figure}
\centering 
   \includegraphics[width=11.5cm]{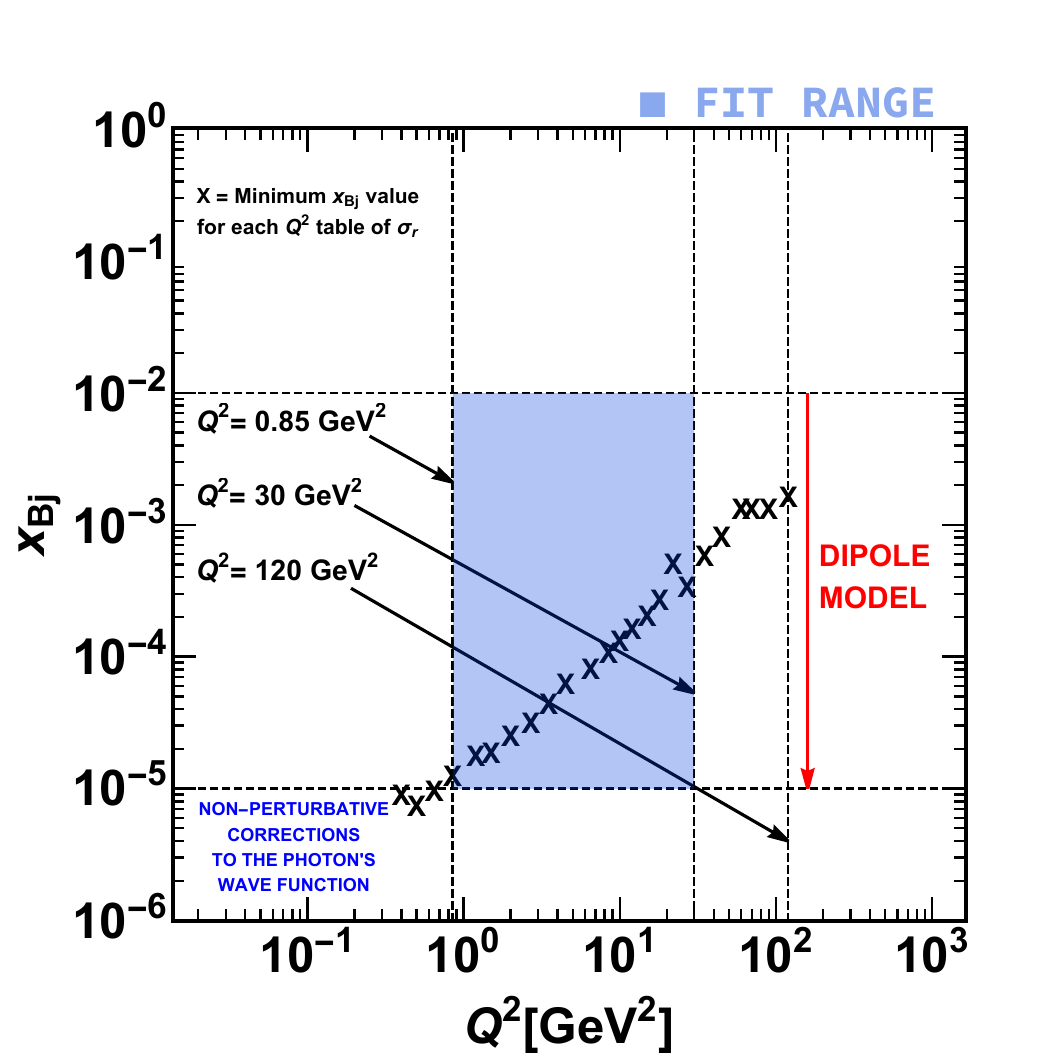}
 \caption{Fit range in the kinematic plane $Q^2$-$x_{\rm{Bj}}$. The diagram illustrates two key regions: the non-perturbative range of the photon's wave function ($Q^2<0.85 \,\rm{GeV}^2$) and the domain where the dipole model holds validity ($x<10^{-2}$). Each marker on the plot represents the minimum $x$ value extracted from experimental data of $\sigma_r$ within specific $Q^2$ intervals, compiled from H1 and ZEUS collaborations ~\cite{HERA1}.}  
\label{FITRANGE}
\end{figure}

\begin{figure}
\centering 
   \includegraphics[width=15.5cm]{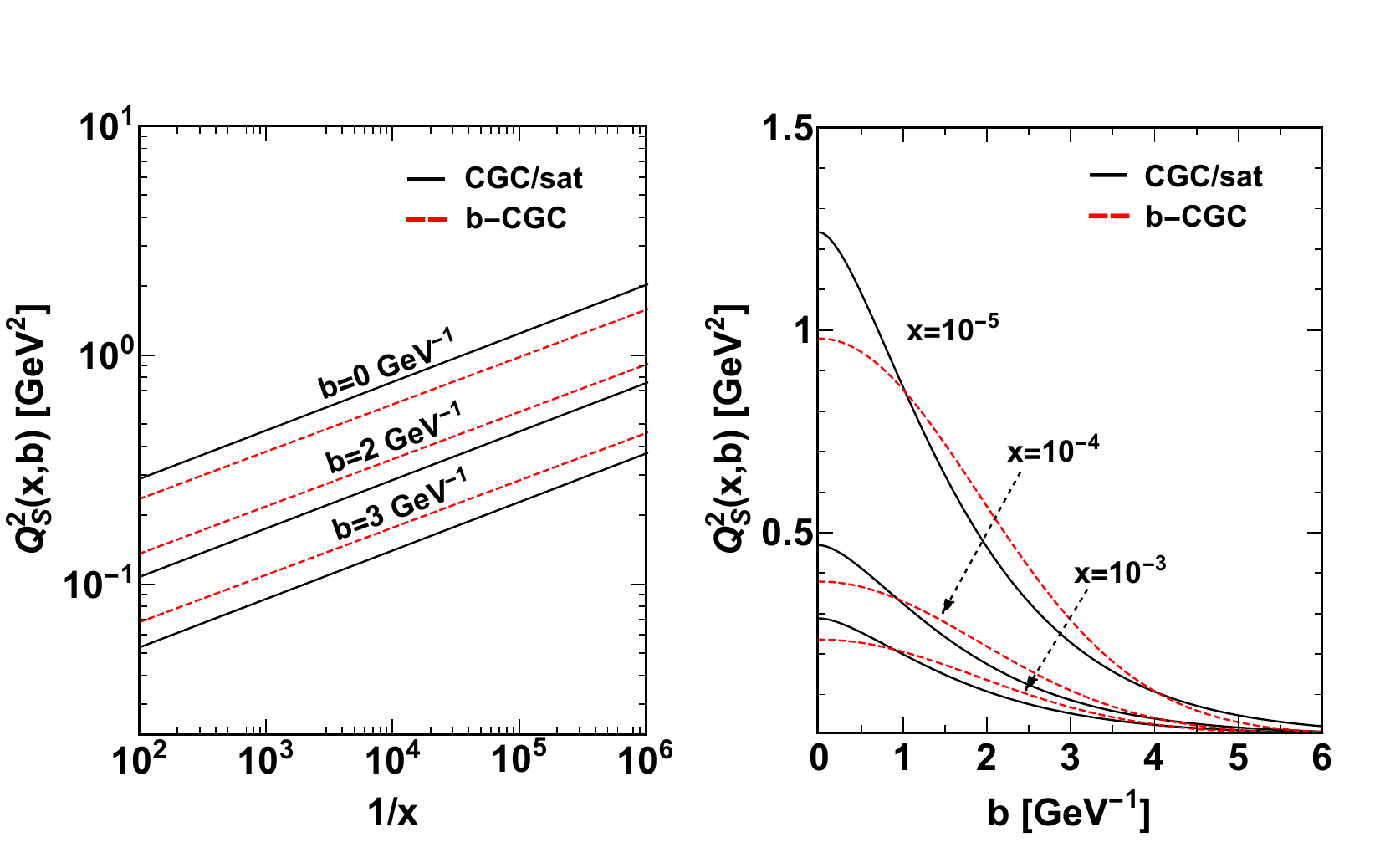}
 \caption{The left panel displays the saturation scale extracted from the CGC/saturation dipole model, illustrating its variation with $1/x$ for different values of $b$. In the right panel, the saturation scale is depicted as a function of the impact parameter $b$ for various fixed values of $x$. Additionally, for comparison, we have included the extracted saturation scale obtained from the b-CGC model \cite{SATMOD17}, utilizing the parameter set with a charm mass of $m_c = 1.27$ GeV for both models.}  
\label{QSxb}
\end{figure}

\section{Numerical results and discussion}

\begin{table}[ht]
{\footnotesize
\begin{tabular}{||l|l|l|l|l|l|l||}
\hline
\hline
\multicolumn{4}{||c|}{Dipole amplitude} & \multicolumn{2}{c|}{Wave function}& \multicolumn{1}{c||}{Minimization} \\
\hline
$\bas$ &  $N_0$ & $Q^2_0$ $(\rm{GeV}^2)$ &$m$ $(\rm{GeV})$ & $m_{u,d,s} $(GeV)&   $m_c $(GeV)& $\ \ \mathrm{\chi^2/d.o.f.}$ \ \ \\ 
\hline
 0.1040 $\pm \, 4.6\times 10^{-4}$ & 0.1512 $\pm \, 3.7\times 10^{-4}$ & 0.824 $\pm \, 3.3\times 10^{-3}$ & 0.5398 $\pm \, 9.6\times 10^{-4}$ & ~~\,$10^{-2}\, \div \,10^{-4}$  & 1.40 & 205.70/166 = 1.239 \\
\hline
 0.1100 $\pm \, 1.8\times 10^{-4}$ & 0.1510  $\pm \, 8.1\times 10^{-4}$ & 0.809 $\pm \, 6.2\times 10^{-3}$ & 0.5412 $\pm \, 1.8\times 10^{-4}$ & ~~\,$10^{-2}\, \div \,10^{-4}$   & 1.27 & 204.73/166 = 1.233  \\
\hline
\hline
\end{tabular}}
\caption{Parameters of the CGC/saturation dipole model: $\bas$, $N_0$,  $Q^2_0$ and  $m$ determined  through fits to the reduced cross-section $\sigma_r$ using the combined H1 and ZEUS data \cite{HERA1} within the range $0.85 \, \rm{GeV}^2<Q^2<30  \, \rm{GeV^2}$ and $x\leq10^{-2}$. Results are presented for fixed light quark masses and two fixed values of the charm quark masses (see the text for details).}
\label{t1}
\end{table}

\begin{figure}
\centering 
   \includegraphics[width=12cm]{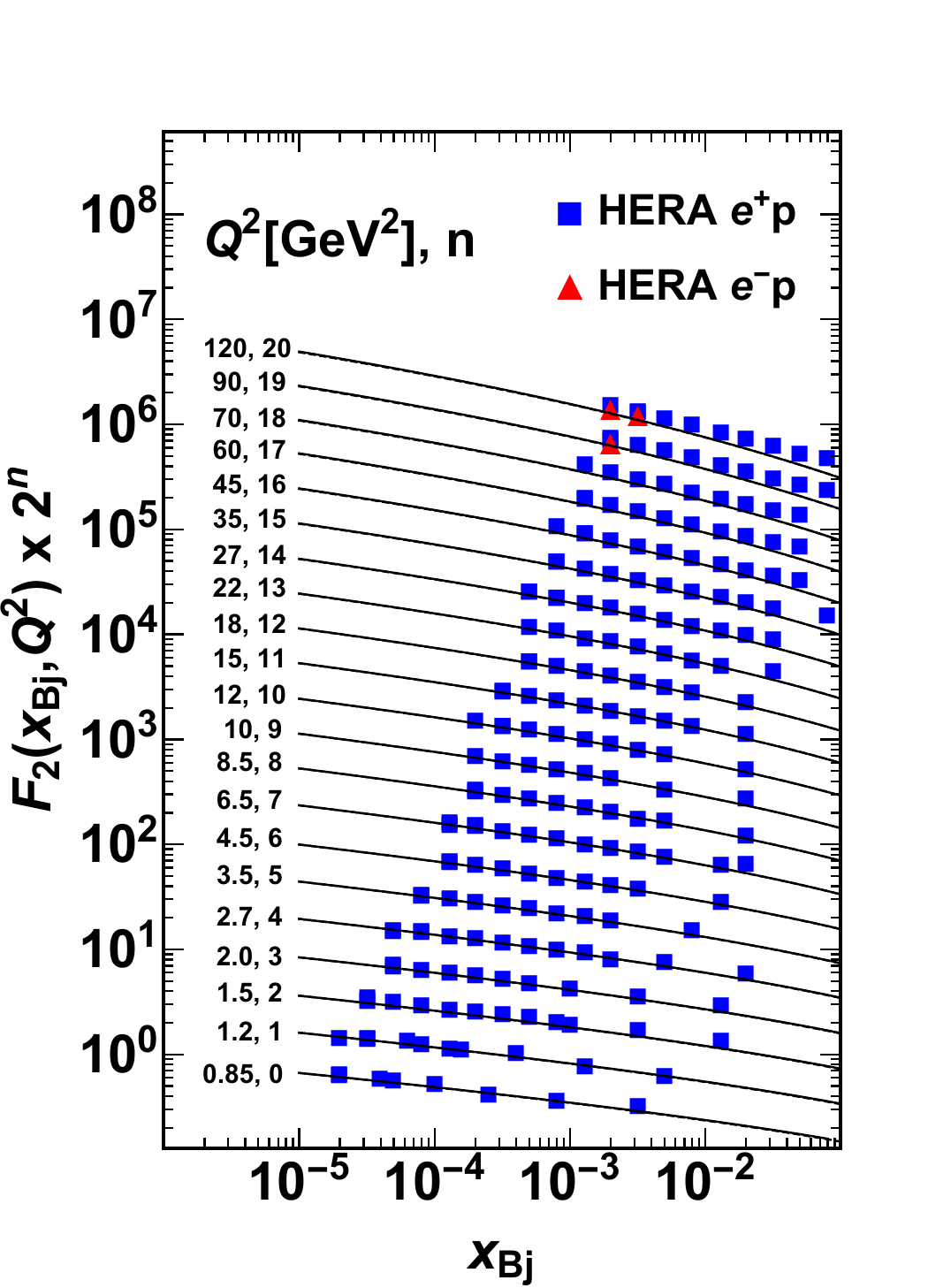}
 \caption{Theoretical estimates for the structure function $F_{2}(x,Q^{2})$ as a function of $x$, for different values of $Q^{2}$. The solid and dashed lines are obtained using two distinct sets of parameters from table I corresponding to charm mass values of $m_c=1.4,\,1.27\,\,\rm{GeV}$, respectively. In order to distinguish each $Q^2$ from the others, both the theoretical results and the experimental data are multiplied by a factor of $2^n$, with the corresponding values of $n$ specified within the plot. The experimental data are from H1 and ZEUS collaborations ~\cite{HERA1}.}  
\label{F2X}
\end{figure}

As we have previously addressed, the model contains four free parameters, which we determine by fitting to experimental data using a $\chi^2$ minimization. We restrict the data used in the fit to reduced inclusive DIS cross-section measurements from the H1 and ZEUS collaborations~\cite{HERA1}. Once the parameters are fixed, we compare the theoretical predictions for $F_{2}$, $F_{L}$, and $F_2^{c\bar{c}}$ with the experimental HERA data. Importantly, the $\chi^2$ calculation incorporates systematic and statistical uncertainties added in quadrature. Our $\chi^2$ calculation incorporates data within a specific range: $0.85\,{\rm GeV}^2 < Q^2 < 30\,{\rm GeV}^2$ and $x \leq 10^{-2}$. This choice balances two key factors. First, the validity of the BK equation, upon which our theoretical framework relies, necessitates small-$x$ values. Therefore, we selected a maximum $x$-value that ensures this condition. Second, we aimed to maximize the data included in the fitting process. The lower $Q^2$ limit stems from non-perturbative corrections applied to the virtual photon wave function. This choice aligns with our fitting results, which exhibit instability for $Q^2$ values below $0.85\,{\rm GeV}^2$. Conversely, the upper $Q^2$ limit was chosen based on fitting stability. While the fit remains stable within the chosen range ($0.85\,{\rm GeV}^2 < Q^2 < 30\,{\rm GeV}^2$), it exhibits instability beyond this threshold. 

Figure~\ref{FITRANGE} depicts the fitting range within the $Q^2$ vs $x_{Bj}$ kinematic plane. The illustration delineates the non-perturbative sector of the photon's wave function ($Q^2<0.85\,\rm{GeV}^2$) as well as the range where the dipole model is applicable ($x<10^{-2}$). The symbol ``X'' represents the smallest $x$-value observed in the experimental data of $\sigma_r$ within each $Q^2$ table from the H1 and ZEUS collaborations~\cite{HERA1} in the specified kinematic region. The vertical lines represent the range of our fit, $Q^2 \in [0.85,30]\,\rm{GeV}^2$, as well as the regions where the observables were predicted. These regions include both the kinematic region defined earlier, as well as regions beyond it, encompassing small values of $Q^2 \leq 0.85\,\rm{GeV}^2$ and large values of $Q^2 \geq 30\,\rm{GeV}^2$.

\begin{figure}
\centering  
   \includegraphics[width=8.5cm]{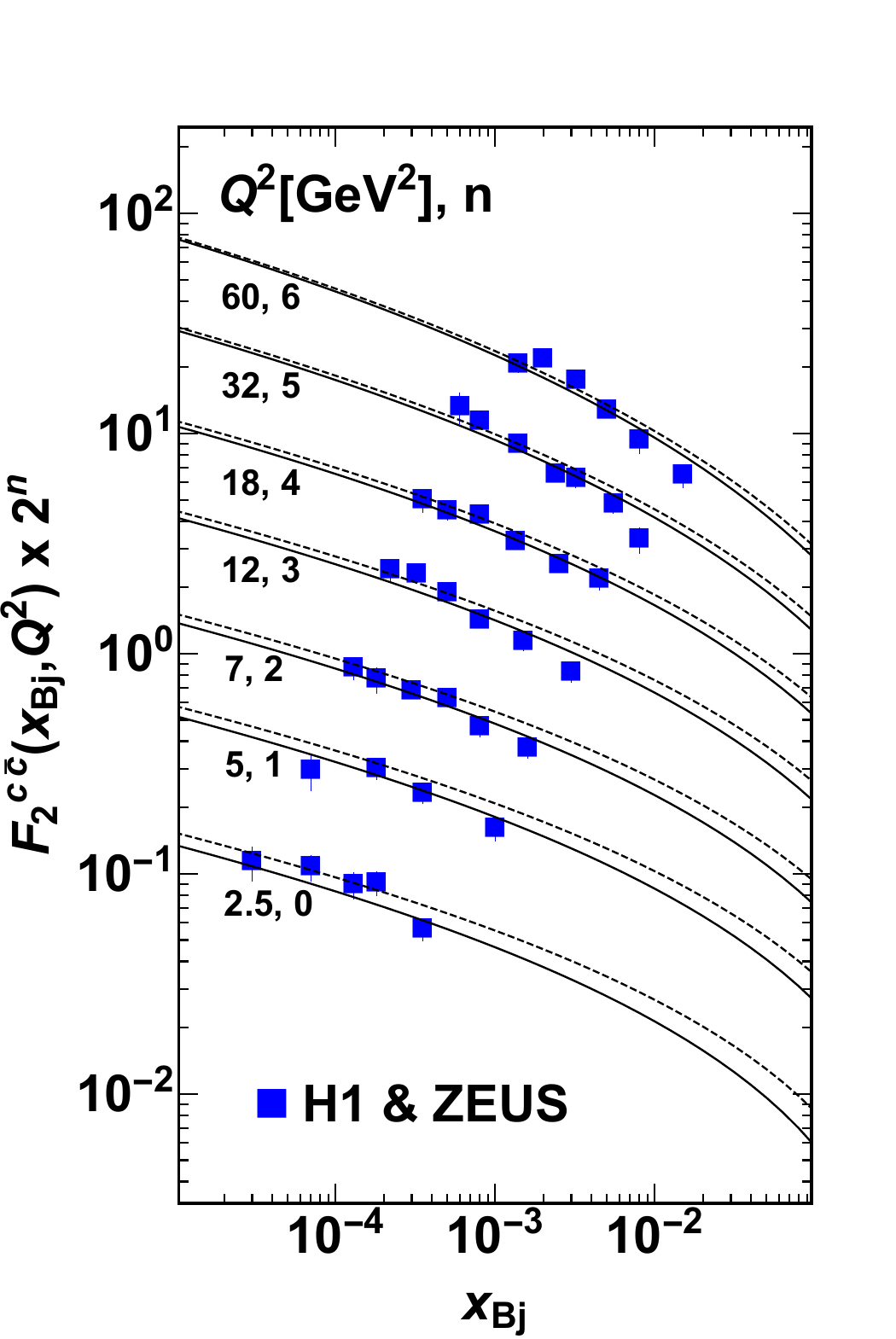}  
\caption{Results for charm structure function $F_{2}^{c\bar{c}}(x,Q^{2})$ as a function of $x$, for different values of $Q^2$. The solid lines correspond to parameters from table I with $m_c=1.4\,\rm{GeV}$, and the dashed lines correspond to parameters with  $m_c=1.27\,\rm{GeV}$. The theoretical estimates as well as the experimental data are multiplied by a factor $2^n$ and the values of $n$ are specified in the plot. The experimental data are from H1 and ZEUS collaboration ~\cite{HERA2}, assuming $\sigma_r^{c\bar{c}}\,\approx\,F_2^{c\bar{c}}$ (see the text for explanation).}    
\label{F2CC}
\end{figure}

\begin{figure}
\centering 
   \includegraphics[width=8.5cm]{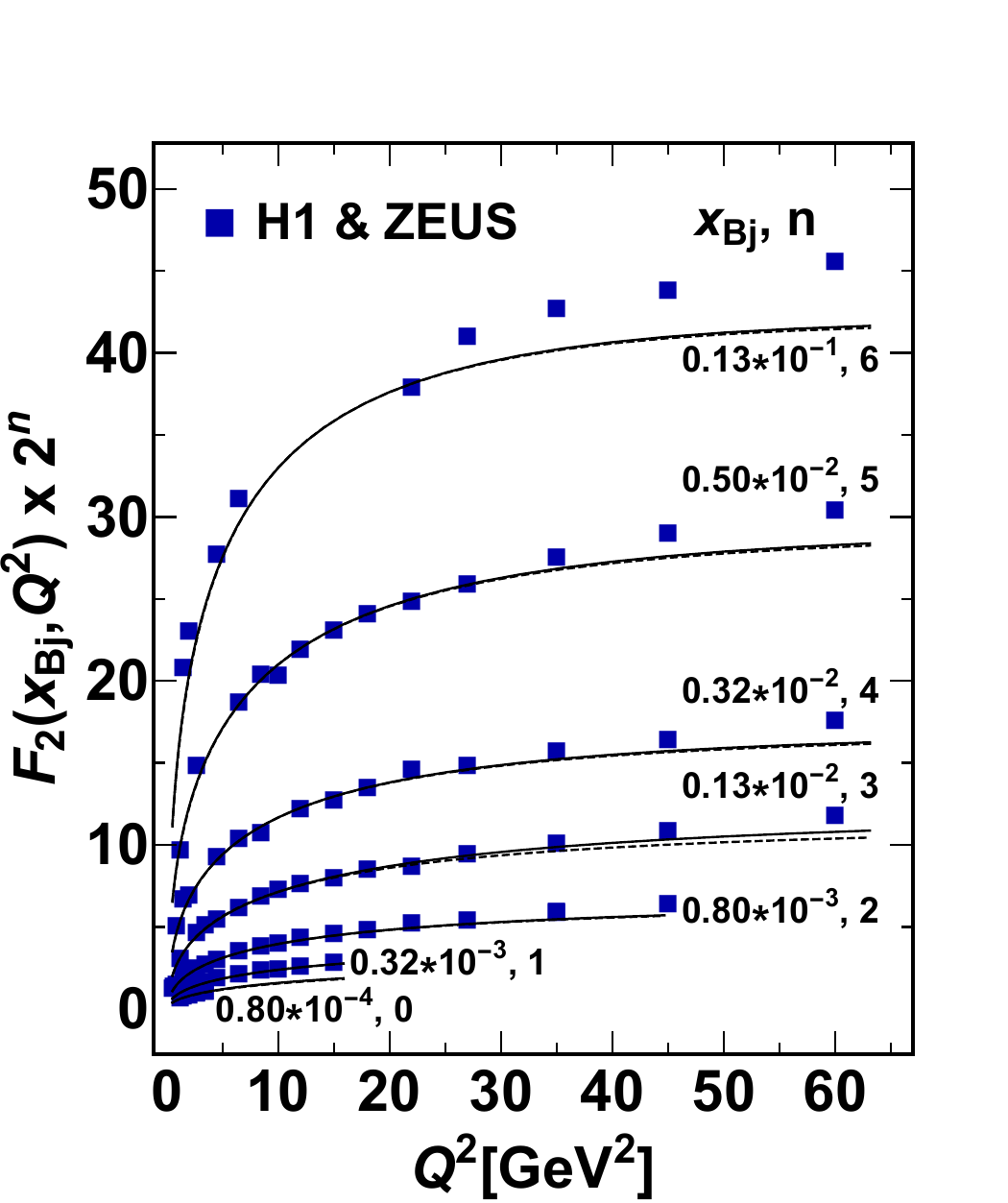}
 \caption{Results for deep inelastic structure function $F_{2}(x,Q^{2})$ as a function of $Q^2$ at fixed values of $x$. The theoretical estimates as well as the experimental data are multiplied by a factor $2^n$ and the values of $x$ and $n$ are specified in the plot. We use the parameters set given in Table I with $m_c=1.4\,\rm{GeV}$ in solid lines and $m_c=1.27\,\rm{GeV}$ in dashed lines. The experimental data are from H1 and ZEUS collaborations ~\cite{HERA1}.} 
\label{F2Q2}
\end{figure}

In our analysis, we conducted fits while varying the light quark masses ($m_u = m_d = m_s$) following  Refs.~\cite{SATMOD16,SATMOD17,SATMOD16A,SATMOD16B}. Initially, we employed a fixed value of $m_{u,d,s} = 0.14\,\rm{GeV}$ for the light quark masses and subsequently explored a wider range from 10$^{-1}$ to 10$^{-4}$ GeV. This variation encompasses a total of five light quark mass configurations. We further explored the impact of the charm quark mass by considering two fixed values: $m_c = 1.4$ GeV and $m_c = 1.27$ GeV. The resulting $\chi^2/d.o.f.$ values for $m_c = 1.4$ GeV were \{1.304, 1.312, 1.239,  1.238, 1.239\}, corresponding to the light quark masses \{0.14, $10^{-1}$, $10^{-2}$, $10^{-3}$, $10^{-4}$\} GeV, respectively. Similarly, for $m_c = 1.27$, the $\chi^2/d.o.f.$ values were \{1.354, 1.292, 1.233, 1.234, 1.233\} for the same set of light quark masses.  The variations in $\chi^2/d.o.f.$ values reflect the stability and quality of the fit as the light quark masses decrease. Specifically, we observed that the model exhibits good fit performance for light quark masses in the range of $10^{-2}$ to $10^{-4}$ GeV. This indicates that the fit becomes more stable and accurate for current values of quark masses of a few MeV. It is important to note that the $\chi^2/d.o.f.$ values included in Table I represent the best fits obtained with the specified light and charm quark masses. These values were chosen because they provided the highest quality fits to the experimental data, as opposed to the less optimal fits that did not reproduce accurately the experimental data. The model is more sensitive to the mass of the light quarks than to the charm mass during the minimization process. Even when including the $\sigma_r^{c\bar{c}}$ data in our $\chi^2$ calculation, it does not significantly influence the parameter values. However, varying the mass value for the light quarks $m_u=0.14$ GeV dramatically changes the parameter values, and the $\chi^2$ value varies between 4\% and 8\% for $m_c = 1.40$ and $m_c = 1.27$, respectively. The resulting values of the free parameters obtained through this fitting process are shown in Table I. One set corresponds to $m_c = 1.40$ GeV, while the other set corresponds to $m_c = 1.27$ GeV.

At the saturation scale, the forward $q\bar{q}$ dipole scattering amplitude $N$ undergoes a rapid increase as $x$ decreases, and the amplitude $N$ becomes significant when non-linear gluon recombination effects turn as prominent as gluon radiation. What differentiates the CGC/saturation dipole model from others is to employ \eq{QS1}  instead of $Q_s\Lb x, b\Rb \,\propto \,\exp\Lb - b^2/B\Rb$ that was used in other saturation models ~\cite{SATMOD5,SATMOD6,SATMOD7,SATMOD8,SATMOD9,SATMOD17,SATMOD12}, which \eq{QS1} provide the correct large $b$ behavior for the scattering amplitude in accord with Froissart theorem~\cite{FROI}. This difference causes the scattering amplitude to increase rapidly in this parametrization, influencing the high-energy predictions significantly. Following Refs. \cite{SATMOD3,SATMOD6,SATMOD16,SATMOD17}, we define the saturation scale $Q_S^2=2/r_S^2$, where $r_S$ is the saturation radius, as a scale where the dipole scattering amplitude has a value $N(x,r_S,b)=(1-\rm{exp}(-1/2))=0.4$\footnote{To ensure the reliability of our results, we incorporate a fixed value of $N(x,r_S,b) = 0.4$ in the formula for comparison with the b-CGC model~\cite{SATMOD17}. It is essential to note that the CGC/saturation dipole model has an additional contribution in the second term of \eq{NZ}, which must be taken into account.}.  It is worth mentioning that the saturation scale $Q_S^2$ differs from the phenomenological parametrization of \eq{QS1}  with different lower subscript $s$. 

In the left panel of \fig{QSxb}, we depict the saturation scale extracted from both the CGC/saturation dipole model and the b-CGC model~\cite{SATMOD17} as a function of $1/x$ for various impact parameter values $b$. It is evident that the saturation scale increases more rapidly for collisions with smaller impact parameters, especially those close to central collisions ($b\approx0$). Additionally, the saturation scale exhibits significant variations across different impact parameter values. This intricate dependence of the saturation scale on the impact parameter highlights its crucial role and needs to be considered for a comprehensive understanding. In the right panel of \fig{QSxb}, which displays the saturation scale as a function of the impact parameter $b$ for several fixed values of $x$ ($x = 10^{-3}, \, 10^{-4}, \, 10^{-5}$), we observe that the saturation scale derived from \eq{QS1} aligns well with the existing HERA data within the specified $x$-range ($x\in[10^{-5},10^{-2}]$). While the b-CGC model's saturation scale accurately reproduces experimental data, it has limitations in satisfying certain theoretical requirements as discussed in the previous section. Conversely, the saturation scale derived from \eq{QS1} fully complies with theoretical considerations. Both scales, within the CGC framework, show similar behaviors, particularly concerning: (i) evolution at small-$x$ and (ii) dependence on the impact parameter $b$. This dependence includes both the effect of varying $b$ values at a fixed $x$ and the evolution of the $b$ dependence itself with fixed $x$ values.

\begin{figure}
\centering 
   \includegraphics[width=16cm]{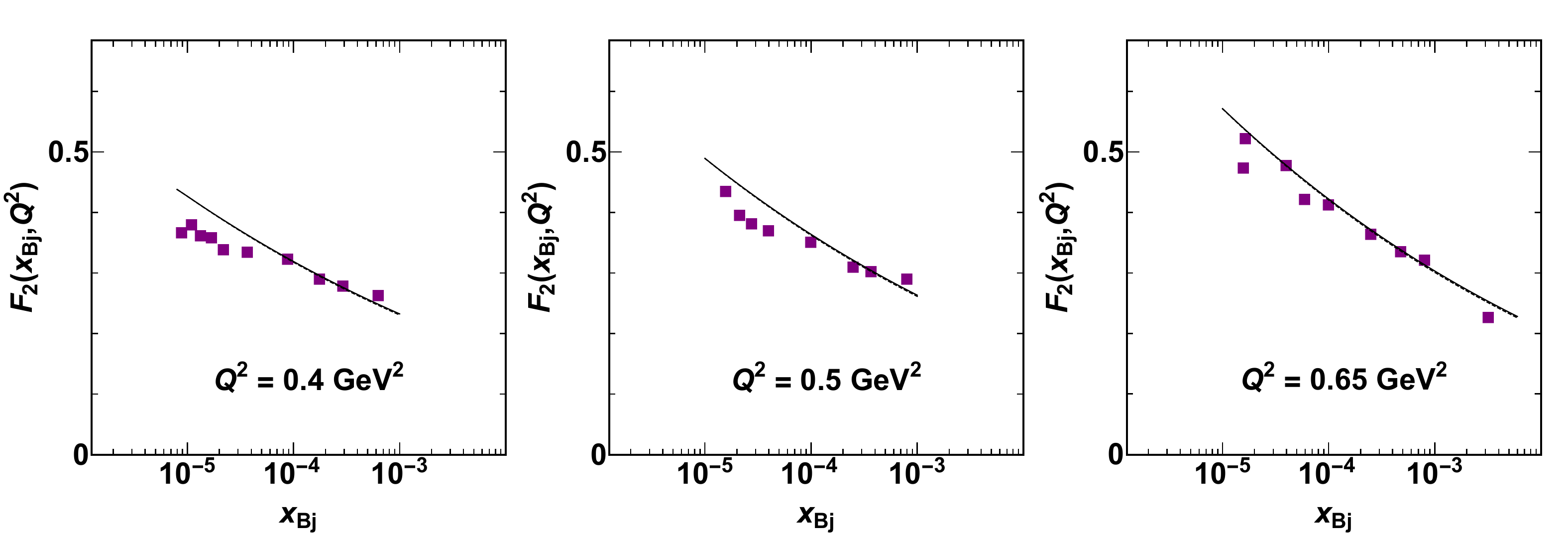}    
\caption{Results for the structure function $F_{2}(x,Q^{2})$ as a function of $x$ at low values of $Q^{2} < 0.85 \ \rm{GeV}^2$. The solid lines correspond to parameter sets given in Table I, with $m_c=1.4\, \rm{GeV}$ and the dashed lines correspond to parameters with $m_c=1.27\, \rm{GeV}$. The experimental data are from H1 and ZEUS collaboration ~\cite{HERA1}.}    
\label{F2LOWQ2}
\end{figure}

\begin{figure}
\centering 
   \includegraphics[width=16cm]{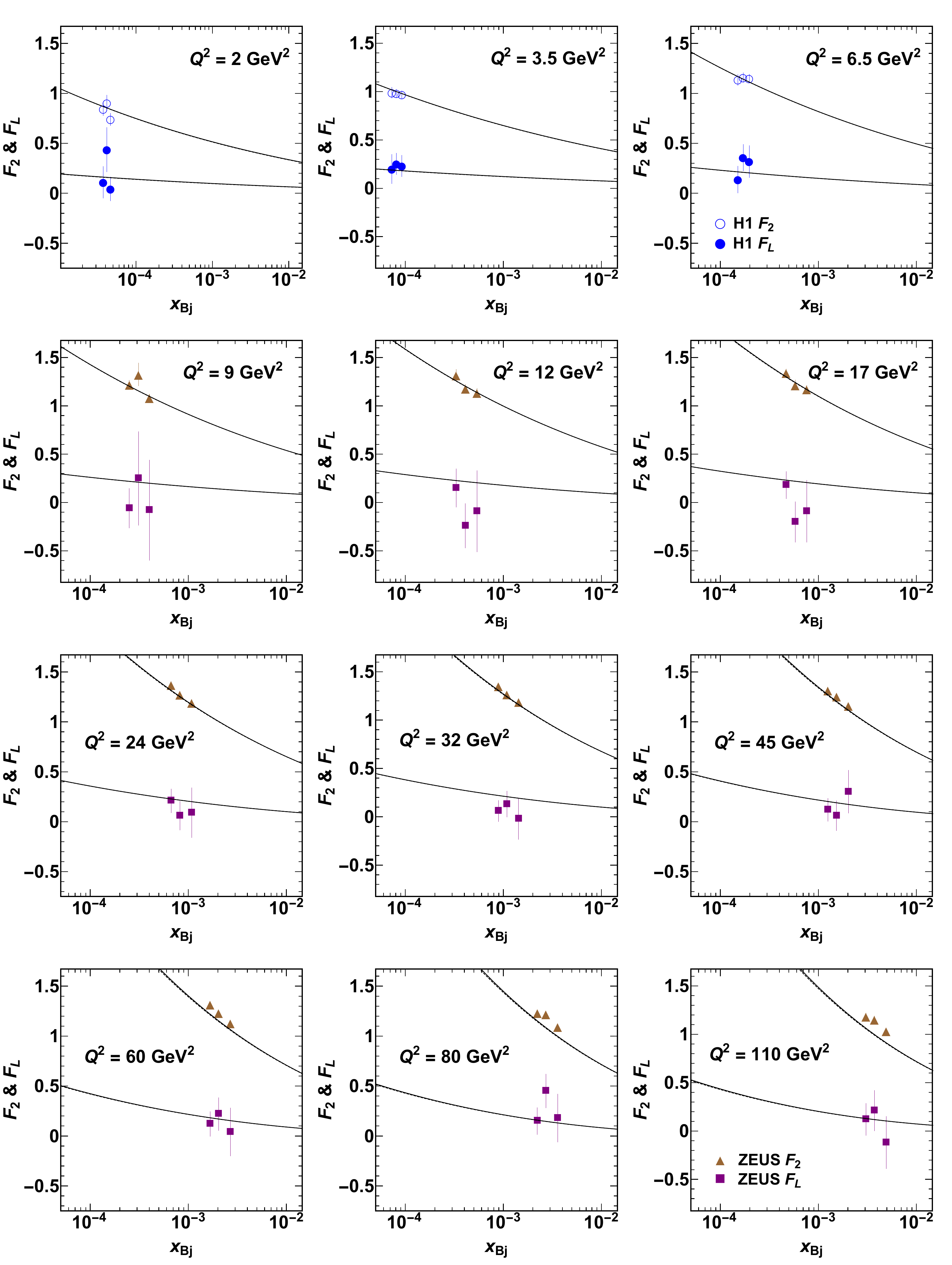}    
\caption{Results for the longitudinal structure function $F_{L}(x,Q^{2})$ and the structure function $F_{2}(x,Q^{2})$, as functions of $x$, for different values of $Q^2$.  The solid and dashed lines are generated using 
parameter sets from table I, which correspond to charm masses of  $m_c=1.4\,\rm{GeV}$ and $m_c=1.27\,\rm{GeV}$, respectively. The experimental data are from ZEUS and H1 collaboration ~\cite{HERAFL2,H1FL}.}    
\label{FL}
\end{figure}

 \begin{table}[ht]
{\footnotesize
\begin{tabular}{||l|l|l||}
\hline
\hline
 \multicolumn{1}{||c|}{Inclusive Processes} & \multicolumn{1}{c||}{$\chi_{d}^2/N_{\rm{points}}$} & \multicolumn{1}{c||}{$\chi_{s}^2/N_{\rm{points}}$ } \\
 \hline
\ \ $F_2$ vs $x_{Bj}$ \ \ & \ \ 1.17\ \ & \ \ 1.18 \ \ \\
\hline
\ \  $F_2^{c\bar{c}}$ vs $x_{Bj}$ \ \ & \ \ 1.15\ \ & \ \ 1.20\ \ \\
\hline
\ \ $F_2$ vs $Q^2$ \ \ & \ \ 1.10\ \ & \ \ 1.10\ \ \\
\hline
\ \ $F_2$ vs $x_{Bj}$ for low $Q^2$\  \ & \ \ 4.87\  \ & \ \ 4.95\ \ \\
\hline
\ \ $F_2$ $\&$ $F_L$ vs $x_{Bj}$ \ \ & \ \ 0.87\ \ & \ \ 0.92\ \ \\
\hline
\hline
\end{tabular}}
\caption{$\chi^2$ values for inclusive processes (Figs. 4-9). Subscripts denote the parameters from Table I used for theoretical calculations for each charm quark mass:  \textit{d} indicates $m_c = 1.27$ GeV (dashed lines), \textit{s}  indicates $m_c = 1.40$ GeV (solid lines).}
\label{t2}
\end{table}

\begin{table}[ht]
{\footnotesize
\begin{tabular}{||l|l|l|l|l|l|l|l|l|l|l|l||}
\hline
\hline
 \multicolumn{1}{||c|}{$\gamma^* p\rightarrow J/\psi \,p$} & \multicolumn{1}{c|}{$\chi_{d}^2$}& \multicolumn{1}{c|}{$\chi_{s}^2$} & \multicolumn{1}{c|}{$\gamma^* p\rightarrow \phi \,p$} & \multicolumn{1}{c|}{$\chi_{d}^2$}& \multicolumn{1}{c|}{$\chi_{s}^2$} & \multicolumn{1}{c|}{$\gamma^* p\rightarrow \rho \,p$} & \multicolumn{1}{c|}{$\chi_{d}^2$}& \multicolumn{1}{c|}{$\chi_{s}^2$} & \multicolumn{1}{c|}{$\gamma^* p\rightarrow \gamma \,p$} & \multicolumn{1}{c|}{$\chi_{d}^2$} & \multicolumn{1}{c||}{$\chi_{s}^2$} \\
\hline
\ \ $\sigma$ vs $Q^2$ \ \ & \ \ 0.33\ \ & \ \ 0.54 \ \  & \ \ $\sigma$ vs $Q^2$ \ \ &\ \ 2.48 \ \ &\ \ 2.62\ \ & \ \ $\sigma$ vs $Q^2$ \ \ & \ \ 0.71 \ \ & \ \ 0.67 \ \ & \ \ $\sigma$ vs $Q^2$ \ \ &\ \ 2.23 \ \ &\ \ 1.62\ \  \\
\hline
\ \  $\sigma$ vs $W$ \ \ & \ \ 0.44\ \ & \ \ 0.61\ \ &\ \  $\sigma$ vs $W$ \ \ & \ \  2.80 \ \ & \ \  2.93 \ \ & \ \ $\sigma$ vs $W$ \ \ & \ \ 0.85\ \ & \ \ 0.73 \ \ & \ \ $\sigma$ vs $W$ \ \ &\ \ 2.38 \ \ &\ \ 1.78\ \ \\
\hline
\ \ $B_D$ vs $Q^2$ \ \ & \ \ 1.21\ \ & \ \ 1.39 \ \ &\ \  $B_D$ vs $Q^2$  \ \  &\ \ 1.65 \ \  &\ \ 1.52 \ \ & \ \ $B_D$ vs $Q^2$ \ \ & \ \ 1.49\ \ & \ \ 1.10\ \ & \ \ $B_D$ vs $Q^2$ \ \ &\ \ 5.93 \ \ &\ \ 6.50 \ \ \\
\hline
\ \ $d\sigma/dt$ vs $|t|$ \ \ & \ \ 1.22\ \ & \ \ 1.34 \ \ &\ \  $d\sigma/dt$ vs $|t|$  \ \  &\ \ 1.58 \ \  &\ \ 1.72 \ \ & \ \ $d\sigma/dt$ vs $|t|$\ \ & \ \ 2.19\ \ & \ \ 2.28 \ \ & \ \ $d\sigma/dt$ vs $|t|$  \ \ &\ \  6.45 \ \ &\ \ 6.99 \ \ \\
\hline
\hline
\end{tabular}}
\caption{$\chi^2$ values for exclusive processes (Figs. 9-14) are shown. Subscripts \textit{d} and \textit{s} correspond to parameter sets (see Table I) used for calculations at different charm quark masses ($m_c = 1.27$ GeV and $m_c = 1.40$ GeV, respectively). Line styles (dashed and solid) indicate the mass values.}
\label{t3}
\end{table}

We can only contrast the model with Refs.~\cite{CLP,CLMP,CLS} since in all other saturation models the assumptions of the saturation scale were made based on Gaussian behavior which contradicts the theoretical information. 
In Ref.~\cite{CLMP}, the $\bas$ value is within the range $\bas\approx 0.14-0.15$, whereas  in Ref.~\cite{CLS} $\bas$ falls in the range $\bas\approx 0.09-0.04$, and in this paper we have found $\bas \, \approx \, 0.10\,\div \, 0.11$.  As we have discussed previously we have obtained $\lambda\,\approx\,0.20$, and this parameter is essential for determining the energy dependence of the saturation scale which is necessary to describe HERA data.  The value of $Q_0^2$ found via $\chi^2$ is smaller than in Ref.~\cite{CLMP} and Ref.~\cite{CLS}. Furthermore, the value of $m$ obtained during the $\chi^2$ calculation is closer to the scale of the gluon mass~\cite{CLP,CLMP}. However, it is important to emphasize a simple distinction between this model and the b-CGC model~\cite{SATMOD17}: the absence of an impact parameter profile for the saturation scale to adjust the slope of the $t$-distribution in diffractive production. Instead, this behavior is governed by the parameter $m$, presenting a greater challenge for this model in accurately reproducing experimental data for diffractive processes.  Alternatively, one could view this as an advantage of the model, as it only requires four parameters instead of five\footnote{The b-CGC model~\cite{SATMOD17} boasts four freely adjustable parameters obtained by fitting reduced sigma data. However, a fifth parameter exists, and unlike the others, it is not freely adjusted. Instead, it was fixed by fitting exclusive $J/\psi$ photoproduction data. This separation prevents a high correlation between $B_{CGC}$ and other DIS model parameters, a correlation that can lead to significant instabilities in calculations.
In contrast to $b$-CGC model, the CGC/saturation dipole model suffers from a strong correlation between its four free parameters. Further research is required to address the high correlation between the four free parameters in the CGC/saturation dipole model. One approach could be to introduce a fifth parameter, similar to the b-CGC model, and set it by fitting exclusive data of a vector meson like $J/\psi$. This approach has the potential to mitigate the correlation issue. Alternatively,  the existing parameter, $m$, could be improved by fitting separately to exclusive photoproduction data.}. 

With the parameters extracted from the $\chi^2$ calculation to the reduced inclusive DIS cross-section $\sigma_r$, as presented in Table I, we now proceed to compute the proton structure function $F_{2}$, the longitudinal structure function $F_{L}$, and the charm structure function $F_2^{c\bar{c}}$. We employ Eqs.~\eqref{F2}, \eqref{FCC}, \eqref{L}, and \eqref{NZ} for these computations and then compare them to the combined HERA data sets. The quality of the fit can be observed in \fig{F2X}. Additionally, we expand our theoretical estimates to high $Q^2$ values, showing a significant agreement with the data. It is important to note that in this specific kinematic region ($Q^2>30\,\rm{GeV}^2$), $F_2^{c\bar{c}}$ was not included in the fitting process, making it a prediction of the CGC/saturation dipole model. As depicted in Figures \ref{F2X}, \ref{F2CC}, \ref{F2Q2} and \ref{FL}  it is clear that the CGC/saturation dipole model results exhibit strong agreement with the data on structure functions, over a wide kinematic range: for $Q^2 \in[0.85,120]\,\rm{GeV}^2$. Additionally, we have included in Table II the $\chi^2/N_{\rm{points}}$ values for each plot of the inclusive processes, covering from Figure \ref{F2X} to Figure \ref{FL}. The model demonstrates evident consistency with the data, particularly regarding the structure functions $F_2$ and $F_2^{c\bar{c}}$, with $\chi^2$ values of 1.17 and 1.13, respectively,  known for their higher precision in experimental data. For the structure functions $F_2$, the agreement with the data remains robust even for $x>10^{-2}$ (a kinematic range beyond the scope of the dipole model, see figure \ref{FITRANGE}) and only begins to weaken as we approach to $x\,\approx\,10^{-1}$. This consistency can also be taken as a non-trivial test of the model's accuracy, stemming from the two critical parameters governing high-energy scattering: the BFKL Pomeron intercept and the energy behavior of the saturation momentum. Consequently, the theoretical results align well with available experimental data. 

When comparing theoretical results with data for the structure-function $F_L$ from the ZEUS and H1 collaborations~\cite{HERAFL2,H1FL}, Fig.~\ref{FL} demonstrates excellent agreement between the model and data ($\chi^2 = 0.87$) within the specified kinematic range (see Fig.~\ref{FITRANGE}) using the two-parameter sets from Table I.  Nevertheless, more accurate measurements of $F_L$ data are required. In Figs. \ref{F2X}, \ref{F2CC}, and  \ref{FL} we extend our theoretical results outside the kinematics of existing data, offering insights for future DIS experiments. \fig{F2CC} illustrates the $F_2^{c\bar{c}}$ data under the assumption that $\sigma_r^{c\bar{c}}\,\approx\,F_2^{c\bar{c}}$. It is important to note that the contribution of $F_L^{c\bar{c}}$ to the reduced cross-section (see \eq{SIGMAR}), arising from the exchange of longitudinally polarized photons, can be approximately or even less than a few percent, making it negligible within the considered kinematic range \footnote{In our calculations, it was unnecessary to introduce an additional normalization factor between the light and charm sectors. For instance, in previous studies~\cite{AAMQS}, the authors attempted this approach but encountered difficulties in accurately computing integrals over the impact parameter of a dipole. As explained on page 5, Section 2.4, they simply replaced these integrals with a constant value $\sigma_0$. However, due to the distinct masses of charm and light quarks (resulting in different dipole sizes), such a substitution is evidently incorrect for all flavors, leading to discrepancies with data, particularly in the charm sector. In an effort to address this discrepancy, the authors introduced separate normalization constants $\sigma_0$ for the charm and light sectors. However, they expressed regret multiple times over this peculiar assumption, acknowledging that it undermines the dipole factorization picture. In contrast, this approach, as demonstrated in Refs.~\cite{SATMOD16,SATMOD17,CLS}, achieves excellent agreement with data in both the light and charm sectors using only leading-order analysis. Consequently, introducing an additional normalization factor is unnecessary in this context.}.

Figures~\ref{F2Q2} and~\ref{F2LOWQ2} warrant special attention. Fig.~\ref{F2Q2} shows the $F_2$  structure function as a function of $Q^2$ at various fixed $x$-values. Notably, we achieve excellent agreement with data for $x<0.013$ ($\chi^2=1.10$) using both charm quark masses ($m_c = 1.4$ GeV and $m_c = 1.27$ GeV). However, for the same $x$-values, the agreement weakens outside the fitting range ($Q^2>30\,\rm{GeV^2}$). Fig.~\ref{F2LOWQ2} demonstrates the description of $F_2$ data at low $Q^2$ values by our fits. Here, the agreement with experimental data is less satisfactory ($\chi^2=4.87$). This discrepancy might be related to the lack of a solid theoretical foundation for the wave functions at low $Q^2$. As suggested in Ref.~\cite{CLS}, incorporating a fixed value of $\bas = 0.2$ in the $\chi^2$ calculation might potentially allow using the perturbative QCD wave function for the virtual photon even at relatively low $Q^2$ values. This approach warrants further investigation.

\begin{figure}
\centering 
  \includegraphics[width=18cm]{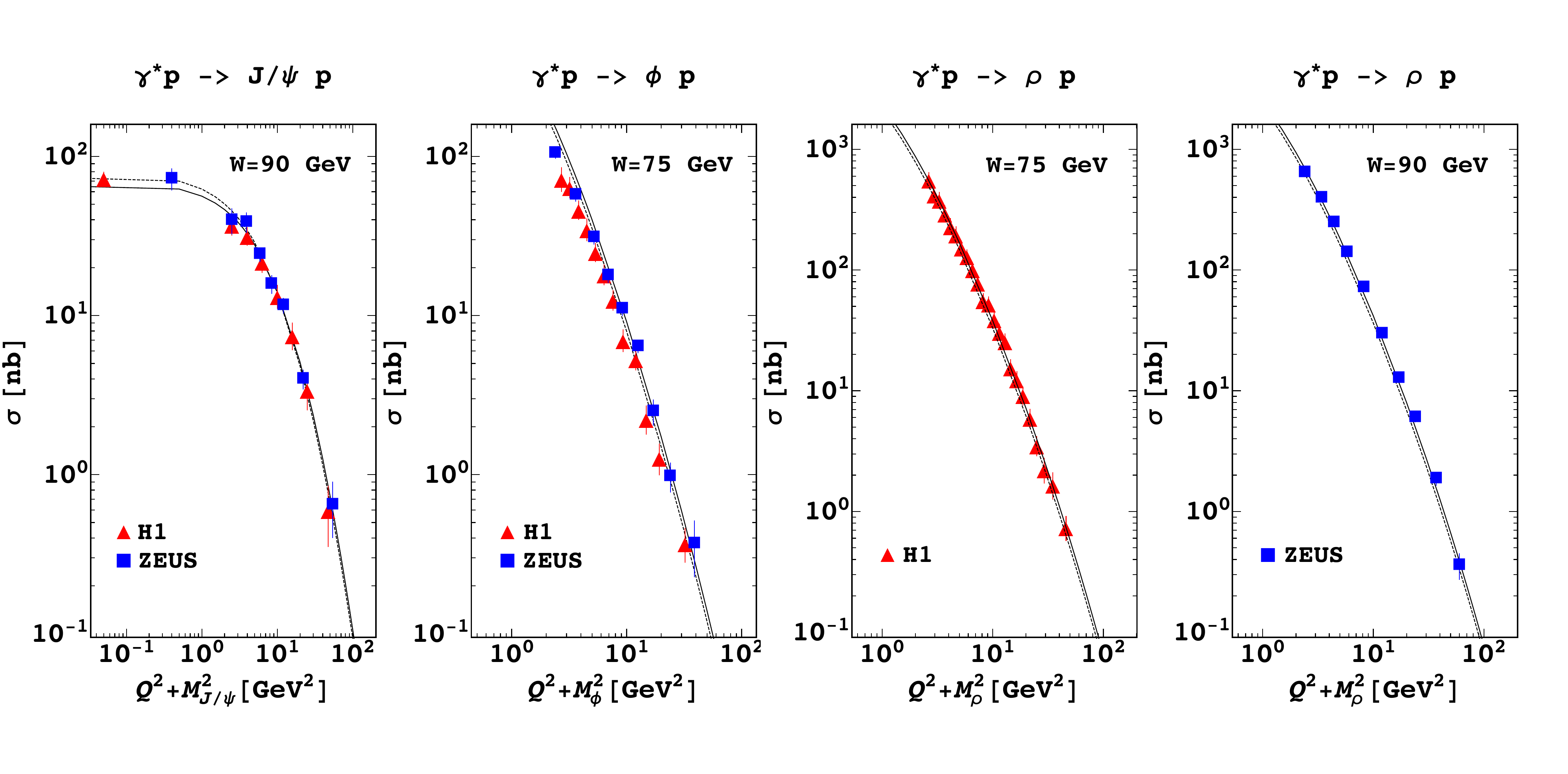}    
\caption{Total vector meson cross-sections $\sigma$ for $J/\psi$, $\phi$ and $\rho$, as a function of $Q^2 + M_{E}^2$ compared to theoretical estimates from CGC/saturation dipole model where solid ($m_c=1.4\,\rm{GeV}$) and dashed ($m_c=1.27\,\rm{GeV}$) lines correspond to the parameters used from table I, respectively. The data are from H1 and ZEUS collaborations ~\cite{EXCL1,EXCL2,EXCL3,EXCL4,EXCL5,EXCL6,EXCL8}.}
\label{DVMP-SIGMA-Q2}
\end{figure}

\begin{figure}
\centering 
   \includegraphics[width=18cm]{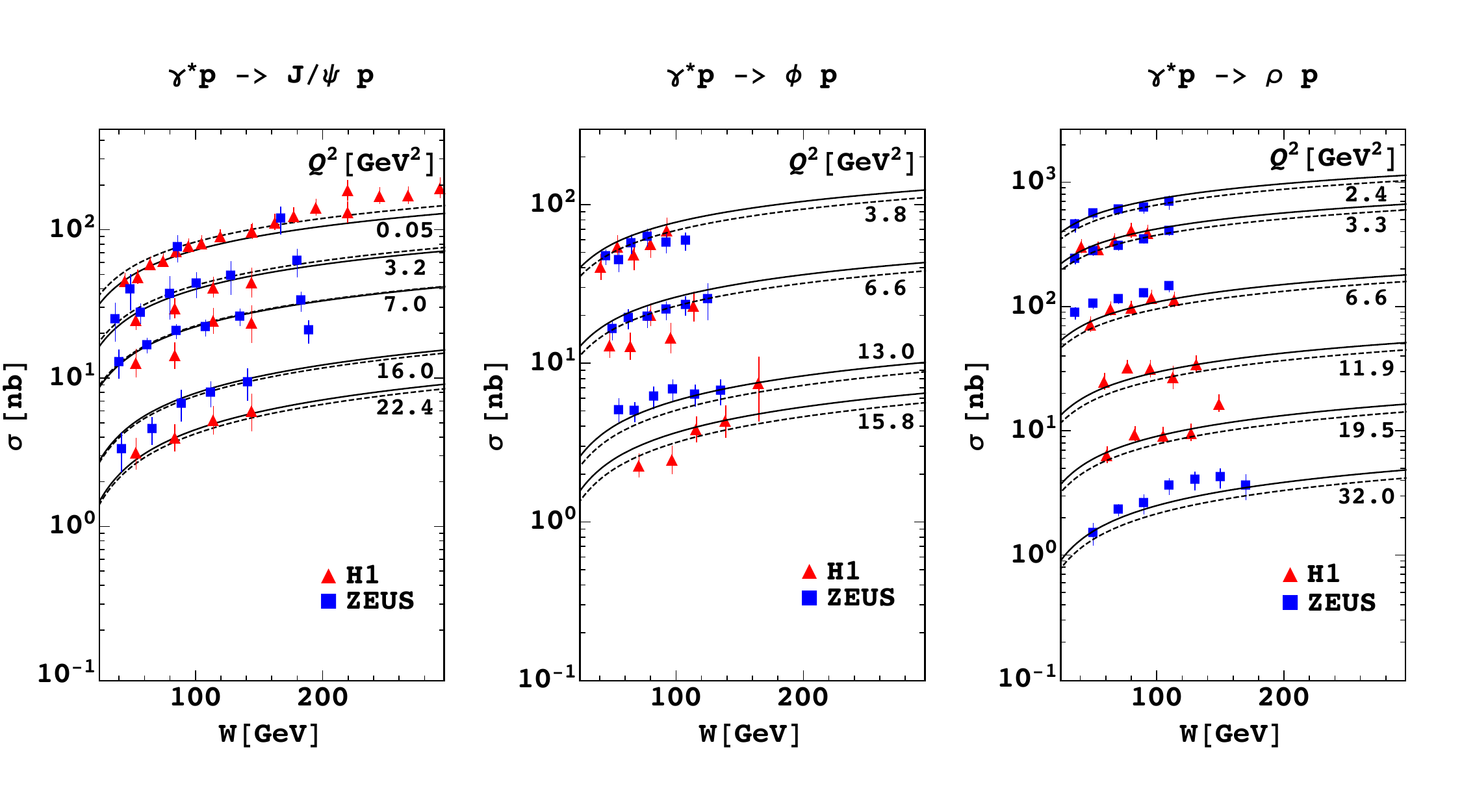}    
\caption{Total vector meson cross-sections $\sigma$ for J/$\psi$, $\phi$ and $\rho$, as a function of $W$. The solid lines
correspond to parameters from table I with  $m_c=1.4\,\rm{GeV}$, and the dashed lines with $m_c=1.27\,\rm{GeV}$. The data are from H1 and ZEUS collaborations ~\cite{EXCL1,EXCL2,EXCL3,EXCL4,EXCL5,EXCL6,EXCL8}.}
\label{DVMP-SIGMA-W}
\end{figure}

\begin{figure}
\centering 
   \includegraphics[width=12.5cm]{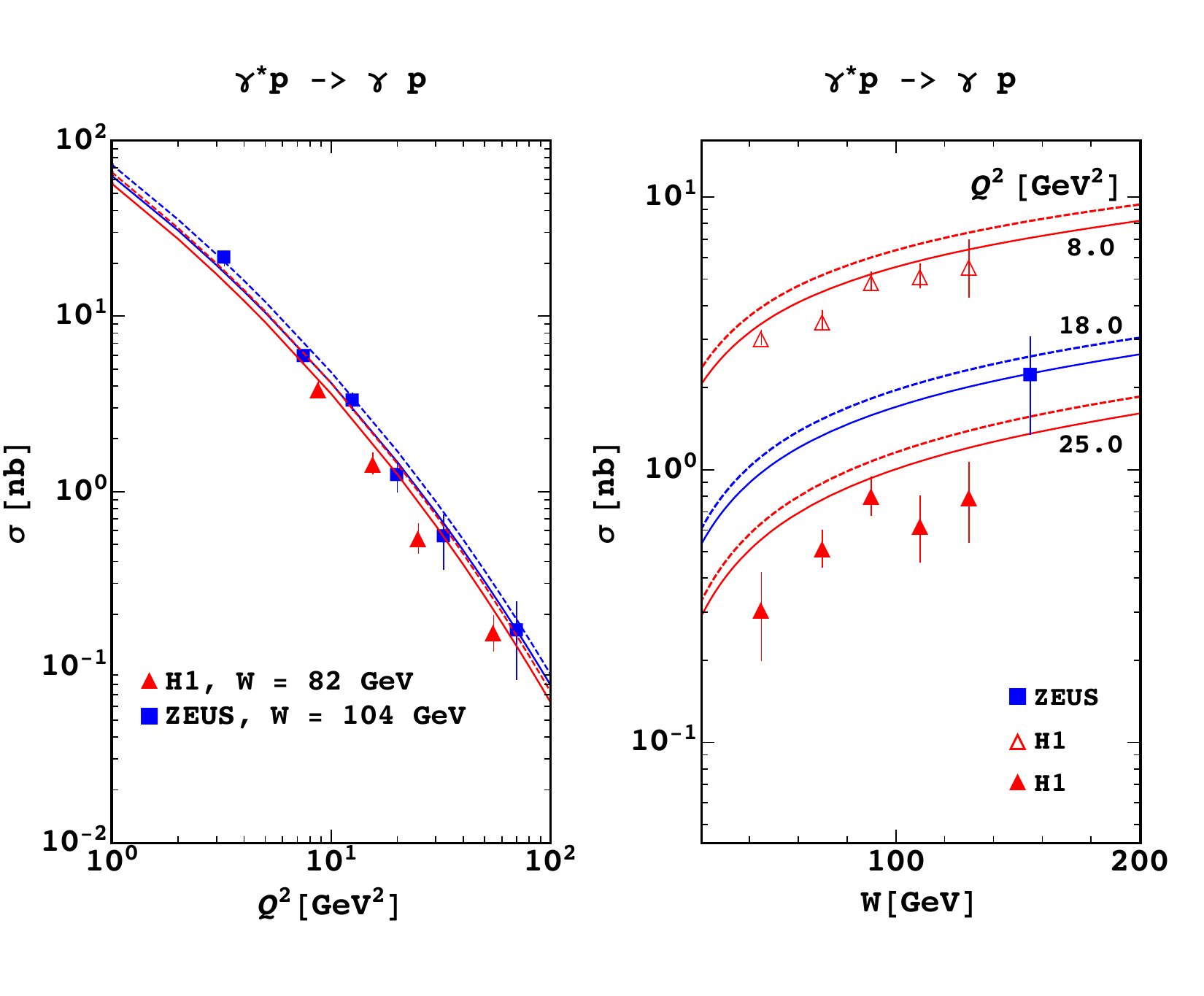}    
\caption{Left: The $Q^2$  dependence of the cross section for DVCS.  Right: The energy dependence of the cross section for DVCS.  For theoretical results, we use the parameter set provided in table I, with $m_c=1.4\,\rm{GeV}$ represented by solid lines and $m_c=1.27\,\rm{GeV}$ depicted by dashed lines. The data are from H1 and ZEUS collaborations ~\cite{EXCL7,EXCL8}.}
\label{DVCS_SIGMA}
\end{figure}

To expand the testing of the CGC/saturation dipole model, we will now examine the exclusive diffractive processes at HERA.  For the total cross-section, in \eq{DSIGMA} we performed the integral over $|t|$ up $1\,\rm{GeV}^2$. In Figs. \ref{DVMP-SIGMA-Q2} and \ref{DVMP-SIGMA-W} we confront the experimental data from H1 and ZEUS with our theoretical predictions for $Q^2$ and $W$-dependence of the vector mesons $J/\psi$, $\phi$ and $\rho$ production in different kinematics. We present our model results computed using two parameter sets from table I, which correspond to charm mass values of $m_c=1.4\,\rm{GeV}$ and $m_c=1.27\,\rm{GeV}$. These results are represented by solid and dashed lines, respectively. One can see that the agreement is excellent, particularly in the case of vector meson $J/\psi$, the wave function (specified in Eqs. \eqref{WFDVMP1}, \eqref{WFDVMP2}, and \eqref{WFDVMP3}) is proficient at replicating experimental data, even at lower values of $Q^2 + M^2_{J/\psi}$ (see \fig{DVMP-SIGMA-Q2}). Notice that both parameter sets give similar estimates, however, for the case of vector meson $J/\psi$ production, its total cross-section exhibits heightened sensitivity to the charm quark mass at low values of $Q^2$ (see \fig{DVMP-SIGMA-W}) which differs from the behavior of lighter mesons $\phi$ and $\rho$. The reason for this lies in the fact that the scale in the integrand of the cross-section is established by $\epsilon_f$ as outlined in \eq{EPSILON}. It is only when virtualities are low, such as for $Q^2 < m_f^2$, that the cross-section starts to exhibit sensitivity to the quark mass.

In Table III, we present the $\chi^2/N_{\rm{points}}$ values for each plot in the figures corresponding to the exclusive processes. Notably, both curves, one parameterized with a charm quark mass of $m_c=1.27$ GeV and the other with $m_c=1.40$ GeV, yield similar $\chi^2$ results. This indicates that the fit provided by each curve is statistically comparable, thereby supporting the consistency and reliability of our findings. Both sets of parameters provide a good description of the experimental data according to the $\chi^2$ values. However, the diffractive production of the light meson $\phi$ deserves special attention. The $\chi^2$ values for $\phi$ meson production are 2.48 for $\sigma\,\, \rm{vs}\,\,Q^2 + M_{\phi}$  and 2.80 for $\sigma\,\,\rm{vs}\,\,W$, suggesting a potential discrepancy between the theoretical predictions and the experimental measurements for this particular meson. This observation aligns with findings from previous studies (Refs. ~\cite{SATMOD16, SATMOD17,CLP}), which also presented a similar situation in accurately predicting $\phi$ meson production data.  Unfortunately, a direct comparison of $\chi^2$ values across studies is not possible because the previous works did not report them.
The CGC/saturation dipole model can be applied to deeply virtual Compton scattering (DVCS) with an accuracy comparable to other discussed reactions. However, limitations arise due to the procedures used to calculate the real part of the amplitude and its skewedness effect can potentially affect the model's ability to describe certain aspects of the data. These limitations are reflected in the $\chi^2$ values obtained when comparing the model with experimental data. While the model provides reasonable agreement for the measured cross-sections, the predicted energy dependence of the cross-section's slope deviates somewhat from the experimental data. Fig. \ref{DVCS_SIGMA} compares the  CGC/saturation dipole model results with experimental data from the H1 and ZEUS collaborations~\cite{EXCL7,EXCL8}. The left panel shows the dependence of the cross-section on $Q^2$ at fixed $W$ values ($W = 82$ GeV and $W = 104$ GeV). This behavior is consistent for both parameter sets from Table I. The right panel presents the $W$-dependence of the cross-section for various fixed $Q^2$ values.

\begin{figure}
\centering 
   \includegraphics[width=18cm]{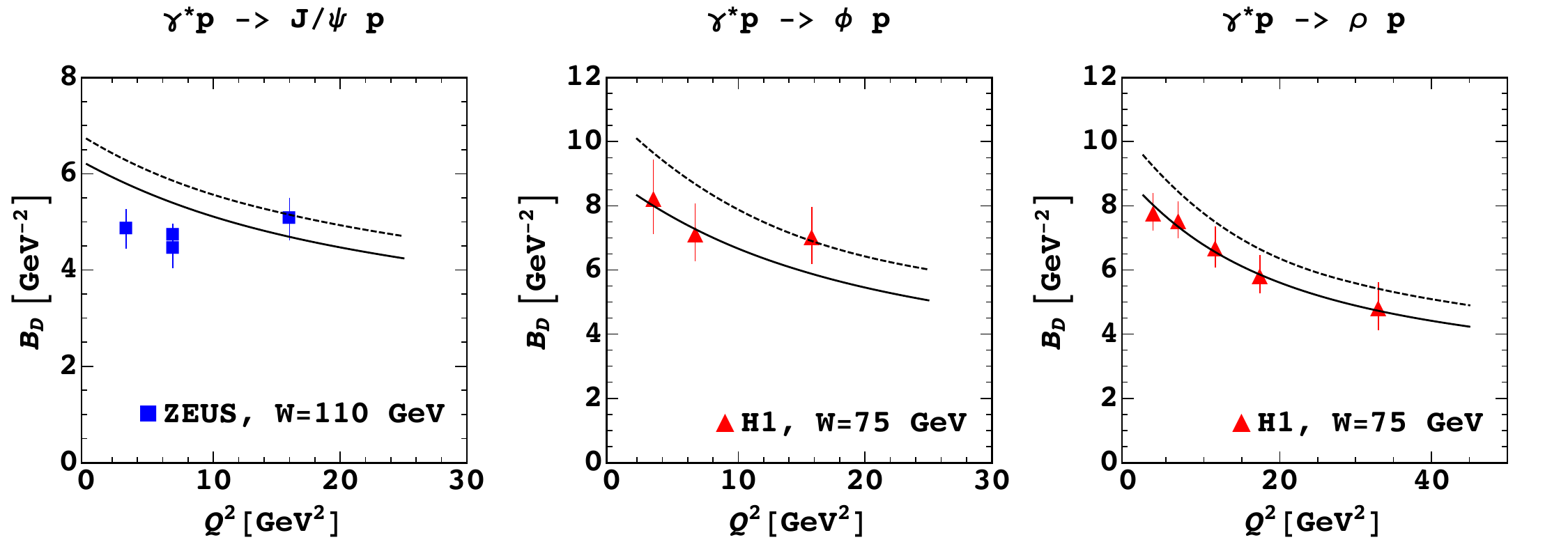}    
\caption{Results for the slope $B_D$ of $t-$distribution of exclusive vector meson electroproduction as a function of $Q^2$. Solid and dashed lines represent calculations using charm quark masses of $m_c = 1.4$ GeV and $m_c = 1.27$ GeV, respectively, from table I. The collection of experimental data are from H1 and ZEUS collaboration ~\cite{EXCL1,EXCL2,EXCL3,EXCL4,EXCL5,EXCL6}.}
\label{BDQ}
\end{figure}

\begin{figure}
\centering 
   \includegraphics[width=11.5cm]{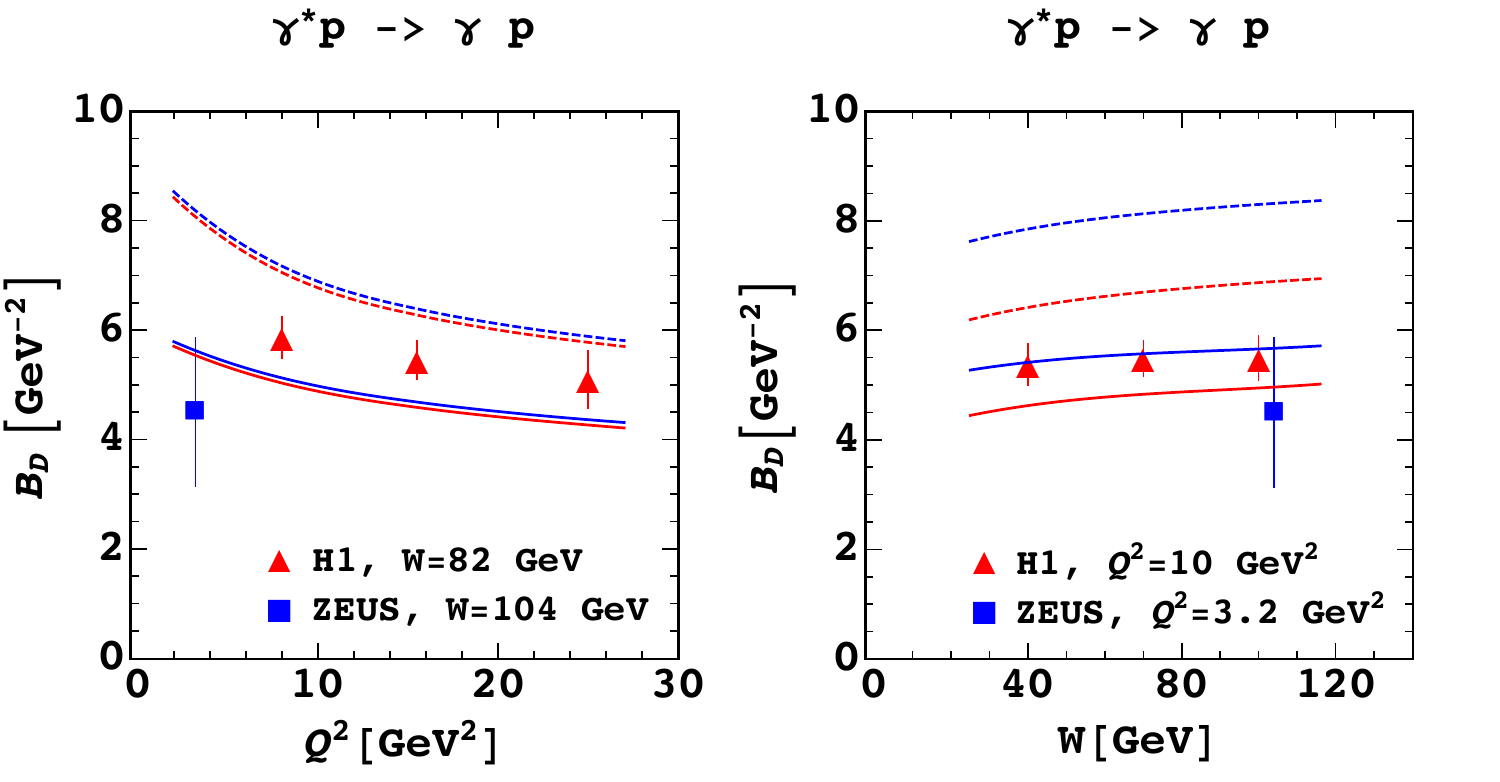}    
\caption{Results for the slope $B_D$ of $t-$distribution of DVCS processes as a function of $Q^2$ and $W$ using parameters from table I. The solid line corresponds to $m_c = 1.4$ GeV, while the dashed line represents $m_c = 1.27$ GeV. The collection of experimental data are from Refs. \cite{EXCL7,EXCL8}.}
\label{BDW}
\end{figure}

\begin{figure}
\centering 
   \includegraphics[width=13.0cm]{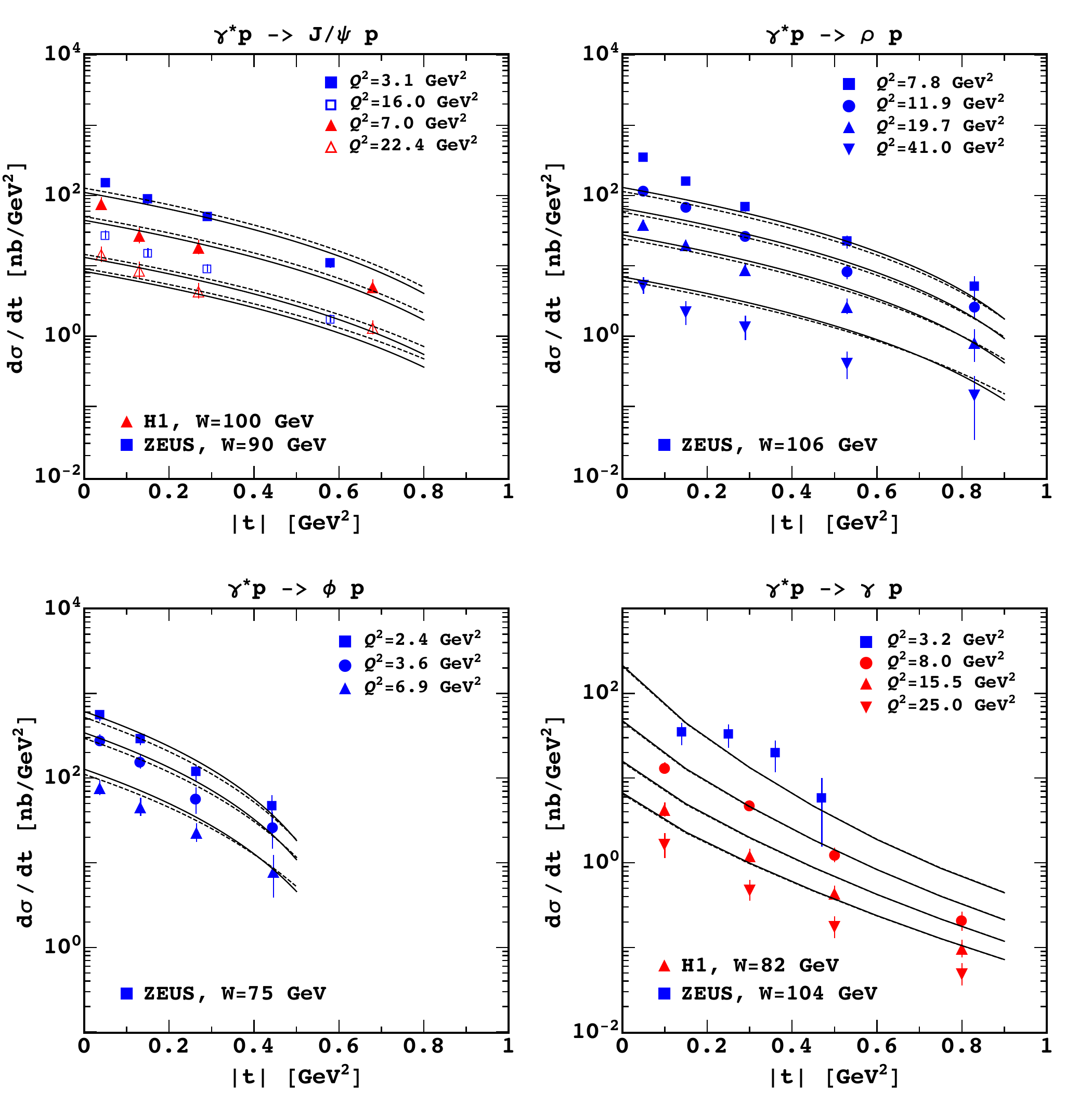}    
\caption{Differential vector meson cross-section for $J/\psi$, $\rho$, $\phi$ and DVCS as a function of |$t$|. The data, corresponding to specific values of $W$ and varying $Q^2$, are compared to theoretical estimates using parameters listed in Table I. The dashed line represents $m_c=1.27$ GeV, while the solid line corresponds to $m_c=1.40$ GeV.  Data are from Refs.\cite{EXCL4, EXCL5, EXCL1, EXCL6, EXCL7}.}
\label{DSIGMA}
\end{figure}

Finally, results for the slope $B_D$ of $t$-distribution of exclusive vector meson production and DVCS are presented in Figs. \ref{BDQ} and \ref{BDW}. One can see that the agreement of the model with the experimental data is reasonable, despite the high $\chi^2$ values reported in table \ref{t3} for DVCS processes. Our analysis reveals that the $B_D$ parameter for lighter vector mesons is higher compared to $J/\psi$ vector mesons produced at the same $Q^2$ values. This observation, consistent with experimental data, indicates a significant difference in the behavior of these mesons under similar kinematic conditions. Additionally, we consider the exclusive diffractive production observed at HERA, as depicted in Figure \ref{DSIGMA}. A key difference between the model presented here and the $b$-CGC model lies in the determination of the parameter $B_{CGC}$, which in the $b$-CGC framework is derived from the $t$-distribution of exclusive diffractive processes at HERA. The distinctive approach of this model compared to the $b$-CGC model, particularly regarding the determination of parameter $m$, offers new insights for understanding the dynamics of exclusive diffractive processes. The observed high correlation among the model's parameters suggests a complex interaction that warrants further investigation. In addition, the proposed model uses a solution of the non-linear BK evolution equation in the saturation regime, rather than an ansatz, ensuring that the theoretical foundation is robust and grounded in established dynamics.  Moreover, the impact parameter distribution resulting in an exponential decrease in the saturation momentum, while satisfying the Froissart bound, is inherently non-perturbative in nature. This non-perturbative origin profoundly influences the $t$-distribution shown in figure \ref{DSIGMA}, particularly in the DVCS process, making it highly sensitive to this nomenclature, which is fundamentally based on phenomenological considerations. The sensitivity of the $t$-distribution to such non-perturbative aspects underscores the importance of carefully considering the impact parameter's role in the analysis, as it directly influences the interpretation of the saturation scale and the overall behavior of the scattering amplitude in this regime. This understanding of the impact parameter's influence on the $t$-distribution also suggests that any comparison with experimental data should account for these non-perturbative effects, which may not be adequately captured by perturbative approaches. Consequently, the model presented here provides a more comprehensive framework for interpreting experimental results in the context of DVCS and vector meson production, especially in regions where conventional perturbative methods may be insufficient. The dispersion in the $t$-distribution for DVCS highlights the need for more detailed studies to explore the potential interpretation of parameter $m$ as the profile of the impact parameter of the saturation scale. Future research should aim to refine these findings, potentially leading to more precise models that can better capture the complexities of exclusive diffractive processes and contribute to the deeper understanding of QCD at high energies.

\section{Conclusions}
In this paper, we confronted the CGC/saturation dipole model of Ref.~\cite{CLMS} with the experimental combined HERA data, determining its parameters through the fitting procedure to the reduced inclusive DIS cross-section $\sigma_r$ at small-$x$. 

Such a model incorporates corrections not only deep within the saturation domain but also in the proximity of the saturation scale, as well as two essential ingredients: \textit{i}) the correct solution to the non-linear BK evolution equation in the saturation region, and \textit{ii}) the impact parameter distribution that results in an exponential decrease of the saturation momentum  as the impact parameter $b$ becomes large and a power-like decrease at high momentum transfers, in accordance with the principles of perturbative QCD. This approach differs from other attempts in that we employ $Q^2_s \propto \,\exp\Lb - m b\Rb$ for large $b$, whereas in other models, Gaussian behavior at large $b$ is assumed, with  $ Q^2_s \propto \,\exp\Lb - \,b^2/B\Rb$. The exponential $b$ dependence of $Q^2_s$ results in the exponential decrease of the scattering amplitude at large $b$, satisfying the Froissart theorem~\cite{FROI}.

The model results are confronted to $F_{2}$, $F_2^{c\bar{c}}$, $F_{L}$, as well as for exclusive diffractive processes such as vector meson production and DVCS with the available data from HERA.  Using four fitting parameters we obtain good overall agreement with the experimental data in the range $Q^2\,\in\,[0.85,120]\,\rm{GeV}^2$ and $x<10^{-2}$. The extracted CGC/saturation dipole model gives the saturation scale for proton $Q_S^2<1\,\rm{GeV}^2$ in accord with other saturation models~\cite{SATMOD16,SATMOD17}. For the structure function $F_2$, the agreement with the data remains robust even for $x>10^{-2}$ and only begins to weaken as $x\,\approx\,10^{-1}$. This consistency serves as a non-trivial test of the model's accuracy to capture the key aspects of high-energy scattering: the BFKL Pomeron intercept and the energy behavior of the saturation momentum. Consequently, the theoretical results align well with available experimental data. However, for $Q^2\, < \,0.85 \,\rm{GeV}^2$ the agreement with experimental data is less satisfactory. This discrepancy might be related to limitations in our current understanding of the wave functions at low $Q^2$. As suggested in Ref.~\cite{CLS}, incorporating a fixed value of $\bas = 0.2$ in the $\chi^2$ calculation might potentially allow using the perturbative QCD wave function for the virtual photon even at relatively low $Q^2$ values. This approach warrants further investigation.

It was illustrated that the CGC/saturation dipole model provides an accurate representation of most aspects in both inclusive DIS and exclusive diffractive data, encompassing the $Q^2$, $W$ , $t$ and $x$ dependence. In Figs. \ref{F2X}, \ref{F2CC}, \ref{FL}, \ref{DVMP-SIGMA-W} and \ref{DVCS_SIGMA}, we extend our theoretical estimates outside the kinematics of existing data, as predictions for forthcoming DIS experiments. The slope $B_D$ of $t$-distribution of exclusive vector meson electroproduction, including $J/\psi$, $\phi$, $\rho$ as well as DVCS can be correctly reproduced, even though the wave functions of the vector mesons and DVCS are quite different. This provides a strong indication of the consistency in the core dynamics and emphasizes the crucial nature of the impact parameter dependence in the saturation scale within the proton. This understanding of the impact parameter's influence on the $t$-distribution highlights the need to account for non-perturbative effects that perturbative methods may not fully capture. The dispersion observed in the $t$-distribution for DVCS emphasizes the necessity for detailed studies to explore the potential interpretation of parameter $m$ as the profile of the impact parameter of the saturation scale. In contrast to b-CGC model, the CGC/saturation dipole model faces a significant challenge due to the high correlation between its four free parameters. To address this, one potential solution could involve introducing a fifth parameter, similar to the b-CGC model, which could be determined by fitting exclusive photoproduction data of a vector meson like $J/\psi$.  This could help mitigate the correlation issue and enhance the model's precision. Alternatively, this approach could also be explored by fitting the parameter $m$, using exclusive photoproduction data. However, further research is needed to clarify this conundrum and make the model more competitive in the market in accurately describing exclusive diffractive processes.

Our results provide a strong guide for finding an approach, based on the Color Glass Condensate/saturation effective theory for high energy QCD, to make reliable predictions from first principles, as well as for upcoming DIS experiments. We believe that the model presented here will be a useful tool to estimate the CGC/saturation effects in a variety of reactions, including the production of dijets in p-p and p-Pb collisions or the structure of the soft Pomeron in CGC. The model takes into account all the theoretical understanding we have regarding deep inelastic processes and, thus, can be used for comparisons at high energy.  In summary, we firmly believe that this model is the only path forward based on solid theoretical ground to extend predictions to higher energy, including those present at the LHC.

\section{Acknowledgements}
We thank our colleagues at UPLA and UTFSM for encouraging discussions, and for the use of the computing infrastructure of the Scientific and Technological Center of Valparaiso, Universidad T\'ecnica Federico Santa Mar\'ia (ANID PIA/APOYO AFB180002). We extend our special thanks to G. Levin and M. Siddikov for their valuable suggestions on this paper. This work was supported by Plan de Fortalecimiento de Universidades Estatales, UPA 19101, CR 18.180, C\'odigo 2390, Universidad de Playa Ancha and ANID Grant No 3230699 (MR), and Grant PIIC $\text{N}^{\rm{o}}$ 029/2023, DPP, Universidad T\'ecnica Federico Santa Mar\'ia (JG).


\begin{thebibliography}{99} \frenchspacing
\bibitem{CLMS}
C.~Contreras, E.~Levin, R.~Meneses and M.~Sanhueza,
{\it ``Non-linear equation in the re-summed next-to-leading order of perturbative QCD: the leading twist approximation,''}
Eur. Phys. J. C \textbf{80} (2020) no.11, 1029
[arXiv:2007.06214 [hep-ph]].

\bibitem{JIMWLK}
~J.~Jalilian-Marian, A.~Kovner, A.~Leonidov and H.~Weigert,
{\it  Phys.\ Rev.}\,  {\bf D59}, 014014 (1999),
[arXiv:hep-ph/9706377];\,\,  {\it Nucl.\ Phys.}\,{\bf B504}, 415
(1997),
[arXiv:hep-ph/9701284]; \,\,\,
J.~Jalilian-Marian, A.~Kovner and H.~Weigert,
  {\it Phys.\ Rev.}  {\bf D59}, 014015 (1999),
  [arXiv:hep-ph/9709432];\,\,\,
 A.~Kovner, J.~G.~Milhano and H.~Weigert,
 {\it  Phys.\ Rev.}  {\bf D62}, 114005 (2000),
  [arXiv:hep-ph/0004014]\,; \,\,\,
E.~Iancu, A.~Leonidov and L.~D.~McLerran,
{\it  Phys.\ Lett.}\,  {\bf B510}, 133 (2001);
[arXiv:hep-ph/0102009];\,\, {\it  Nucl.\ Phys.}\,  {\bf A692}, 583
(2001),
[arXiv:hep-ph/0011241];\,\,\,
E.~Ferreiro, E.~Iancu, A.~Leonidov and L.~McLerran,
 {\it  Nucl.\ Phys.}\  {\bf A703}, 489 (2002),
  [arXiv:hep-ph/0109115];\,\,\,
H.~Weigert,
{\it  Nucl.\ Phys.}  {\bf A703}, 823 (2002),
[arXiv:hep-ph/0004044].

\bibitem{BK}
I.~Balitsky,
[arXiv:hep-ph/9509348];\,\,
 Phys.\ Rev. {\bf D60}, 014020 (1999)
[arXiv:hep-ph/9812311];\,\,\,\,
Y.~V.~Kovchegov,
 Phys.\ Rev.  {\bf D60}, 034008  (1999),
[arXiv:hep-ph/9901281].

\bibitem{SALAM}
G.~P.~Salam,
  {\it ``A Resummation of large subleading corrections at small x,''}
  JHEP {\bf 9807} (1998) 019
  [hep-ph/9806482];
\bibitem{SALAM1}
   M.~Ciafaloni, D.~Colferai and G.~P.~Salam,
  {\it ``Renormalization group improved small $ x$  equation,''}
  Phys.\ Rev.\ D {\bf 60} (1999) 114036
  [hep-ph/9905566].
  
\bibitem{SALAM2}
 M.~Ciafaloni, D.~Colferai, G.~P.~Salam and A.~M.~Stasto,
  {\it ``Renormalization group improved small $x$ Green's function,''}
  Phys.\ Rev.\ D {\bf 68} (2003) 114003,
  [hep-ph/0307188].

\bibitem{DIMST}
B.~Duclou\'e, E.~Iancu, A.~H.~Mueller, G.~Soyez and D.~N.~Triantafyllopoulos,
JHEP \textbf{04}, 081 (2019)
[arXiv:1902.06637 [hep-ph]].

\bibitem{BFKL}
V.~S. Fadin, E.~A. Kuraev and L.~N. Lipatov,
{\it ``On the pomeranchuk singularity in asymptotically free theories"},
\newblock Phys. Lett. {\bf B60}, 50 (1975);\,\,\,
E.~A. Kuraev, L.~N. Lipatov and V.~S. Fadin,
{\it``The Pomeranchuk Singularity in Nonabelian Gauge Theories"}
\newblock Sov. Phys. JETP {\bf 45}, 199 (1977),
\newblock [Zh. Eksp. Teor. Fiz.72,377(1977)];\,\,\,
I.~I. Balitsky and L.~N. Lipatov,{\it ``The Pomeranchuk Singularity in Quantum Chromodynamics,''}
\newblock Sov. J. Nucl. Phys. {\bf 28}, 822 (1978),
\newblock [Yad. Fiz.28,1597(1978)].


\bibitem{LIP}
 L.~N.~Lipatov,
  {\it ``Small x physics in perturbative QCD,''}
  Phys.\ Rept.\  {\bf 286}, 131 (1997)
  [hep-ph/9610276];\,\,\,{\it ``The Bare Pomeron in Quantum Chromodynamics,''}
  Sov.\ Phys.\ JETP {\bf 63}, 904 (1986)
  [Zh.\ Eksp.\ Teor.\ Fiz.\  {\bf 90}, 1536 (1986)].

\bibitem{LETU}
E.~Levin and K.~Tuchin,
  {\it ``Solution to the evolution equation for high parton density QCD,''}
  Nucl.\ Phys.\   {\bf B573}, 833 (2000)
  [hep-ph/9908317];\,\,\,
{\it ``New scaling at high-energy DIS,''}
  Nucl.\ Phys.\  {\bf A691}, 779 (2001)
  [hep-ph/0012167]; {\it ``Nonlinear evolution and saturation for heavy nuclei in DIS,''}
   {\bf 693}, 787 (2001)
  [hep-ph/0101275].  


\bibitem{CLMP}
C.~Contreras, E.~Levin, R.~Meneses and I.~Potashnikova,
{\it``CGC/saturation approach: a new impact-parameter dependent model in the next-to-leading order of perturbative QCD,''}
Phys. Rev. D \textbf{94}, no.11, 114028 (2016)
[arXiv:1607.00832 [hep-ph]].


\bibitem{XCWZ}
W.~Xiang, Y.~Cai, M.~Wang and D.~Zhou,
{\it ``Rare fluctuations of the $S$-matrix at NLO in QCD,''}
Phys. Rev.   \textbf{D99} (2019) no.9, 096026
[arXiv:1812.10739 [hep-ph]].  


\bibitem{NLOBK0}
I.~Balitsky,
  {\it ``Quark contribution to the small-x evolution of color dipole,''}
  Phys.\ Rev.\ D {\bf 75} (2007) 014001,
  [hep-ph/0609105].
  \bibitem{NLOBK01}
 Y.~V.~Kovchegov and H.~Weigert,
  {\it ``Triumvirate of Running Couplings in Small-x Evolution,''}
  Nucl.\ Phys.\   {\bf A784} (2007) 188,
  [hep-ph/0609090].
  
  \bibitem{NLOBK1}
I.~Balitsky and G.~A. Chirilli, {\it ``Next-to-leading order evolution of color
  dipoles"}, \href{http://dx.doi.org/10.1103/PhysRevD.77.014019}
  Phys. Rev. {\bf D77} (2008)  014019 [arXiv:0710.4330 [hep-ph]].

\bibitem{NLOBK2}
I.~Balitsky and G.~A. Chirilli, {\it ``Rapidity evolution of Wilson lines at the
  next-to-leading order"},
  \href{http://dx.doi.org/10.1103/PhysRevD.88.111501} Phys. Rev. {\bf D88}
  (2013)  111501,
\href{http://arxiv.org/abs/1309.7644}[ arXiv:1309.7644 [hep-ph]].

\bibitem{JIMWLKNLO1}
A.~Kovner, M.~Lublinsky, and Y.~Mulian, {\it ``Jalilian-Marian, Iancu, McLerran,
  Weigert, Leonidov, Kovner evolution at next to leading order"},
  \href{http://dx.doi.org/10.1103/PhysRevD.89.061704} Phys. Rev. {\bf D89}
  (2014) no.~6, 061704,
\href{http://arxiv.org/abs/1310.0378}[ arXiv:1310.0378 [hep-ph]].

\bibitem{JIMWLKNLO2}
A.~Kovner, M.~Lublinsky, and Y.~Mulian, {\it ``NLO JIMWLK evolution unabridged"},
  \href{http://dx.doi.org/10.1007/JHEP08(2014)114} JHEP {\bf 08} (2014)
  114,
\href{http://arxiv.org/abs/1405.0418}[ arXiv:1405.0418 [hep-ph]].

\bibitem{JIMWLKNLO3}
M.~Lublinsky and Y.~Mulian, {\it ``High Energy QCD at NLO: from light-cone wave
  function to JIMWLK evolution"},
  \href{http://dx.doi.org/10.1007/JHEP05(2017)097} JHEP {\bf 05} (2017)
  097,
\href{http://arxiv.org/abs/1610.03453}[arXiv:1610.03453 [hep-ph]].


\bibitem{KOLEB}
Yuri V Kovchegov and Eugene Levin, {\it `` Quantum Choromodynamics at High Energies"}, Cambridge Monographs on Particle Physics, Nuclear Physics and Cosmology, Cambridge University Press, 2012 .


\bibitem{FROI}
M.~Froissart,
{\it Phys.\, Rev.} \,  {\bf 123} (1961) 1053; \\
~A. ~Martin, {``Scattering Theory: Unitarity, Analitysity and Crossing."}
Lecture Notes in Physics, Springer-Verlag,  Berlin-Heidelberg-New-York,
1969.  

   \bibitem{KW}
 A.~Kovner and U.~A.~Wiedemann,
  Phys.\ Rev.\ D {\bf 66}, 051502 (2002)
  [hep-ph/0112140];\,\,
  Phys.\ Rev.\ D {\bf 66}, 034031 (2002)
  [hep-ph/0204277];\,\,
  Phys.\ Lett.\ B {\bf 551}, 311 (2003)
  [hep-ph/0207335].

  \bibitem{FIIM} 
  E.~Ferreiro, E.~Iancu, K.~Itakura and L.~McLerran,
  {\it ``Froissart bound from gluon saturation,''}
  Nucl.\ Phys.\   {\bf A710}, 373 (2002)
  [hep-ph/0206241].

\bibitem{GS}
J.~Bartels, E.~Levin,
  Nucl.\ Phys.\  {\bf B387 } (1992)  617-637;\,\,
  L.~McLerran, M.~Praszalowicz,
  Acta Phys.\ Polon.\  {\bf B42 } (2011)  99,
  [arXiv:1011.3403 [hep-ph]]  {\bf B41 } (2010)  1917-1926,
  [arXiv:1006.4293 [hep-ph]];\,\,\,M.~Praszalowicz,
  Acta Phys.\ Polon.\ B {\bf 42} (2011) 1557
  [arXiv:1104.1777 [hep-ph]];\,\,
  M.~Praszalowicz and T.~Stebel,
  JHEP {\bf 1303} (2013) 090
  [arXiv:1211.5305 [hep-ph]];\,\,
  L.~McLerran, M.~Praszalowicz and B.~Schenke,
  Nucl.\ Phys.\ A {\bf 916} (2013) 210
  [arXiv:1306.2350 [hep-ph]];\,\,
  M.~Praszalowicz,
  Phys.\ Lett.\ B {\bf 727} (2013) 461
  [arXiv:1308.5911 [hep-ph]];\,\,
  L.~McLerran and M.~Praszalowicz,
  Phys.\ Lett.\ B {\bf 741} (2015) 246
  [arXiv:1407.6687 [hep-ph]].

\bibitem{SGBK}
 A.~M.~Stasto, K.~J.~Golec-Biernat, J.~Kwiecinski,
  {\it ``Geometric scaling for the total gamma* p cross-section in the low x region,''}
  Phys.\ Rev.\ Lett.\  {\bf 86 } (2001)  596-599,
  [hep-ph/0007192].  

\bibitem{BKL}
S.~Bondarenko, M.~Kozlov and E.~Levin,
{\it ``QCD saturation in the semi-classical approach,''}
Nucl. Phys.   \textbf{A727} (2003), 139-178
[arXiv:hep-ph/0305150 [hep-ph]].

 \bibitem{SATMOD0}
     K.~J.~Golec-Biernat and M.~Wusthoff,
  {\it `Saturation in diffractive deep inelastic scattering,''}
  Phys.\ Rev.\  {\bf D60} (1999) 114023
  [hep-ph/9903358];\,\, {\it ``Saturation effects in deep inelastic scattering at low Q**2 and its implications on diffraction,''}
  Phys.\ Rev.\  {\bf D59} (1998) 014017;\,\,
  [hep-ph/9807513].
\bibitem{SATMOD01}
J.~Berger and A.~M.~Stasto,
``Small x nonlinear evolution with impact parameter and the structure function data,''
Phys. Rev. D \textbf{84} (2011), 094022
[arXiv:1106.5740 [hep-ph]].
  
 \bibitem{SATMOD1}
    J.~Bartels, K.~J.~Golec-Biernat and H.~Kowalski,
  {\it ``A modification of the saturation model: DGLAP evolution,''}
  Phys.\ Rev.\  {\bf D66} (2002) 014001
  [hep-ph/0203258].  

    \bibitem{SATMOD2}
   H.~Kowalski and D.~Teaney,
  {\it ``An Impact parameter dipole saturation model,''}
  Phys.\ Rev.\  {\bf D68} (2003) 114005
  [hep-ph/0304189].

    \bibitem{IIM}
   E.~Iancu, K.~Itakura and S.~Munier,
  {\it ``Saturation and BFKL dynamics in the HERA data at small x,''}
  Phys.\ Lett.\  {\bf B590} (2004) 199
  [hep-ph/0310338].
  
\bibitem{SATMOD3}
    H.~Kowalski, L.~Motyka and G.~Watt,
{\it ``Exclusive diffractive processes at HERA within the dipole picture,''}
  Phys.\ Rev.\  {\bf D74} (2006) 074016
  [hep-ph/0606272].
  \bibitem{SATMOD4}
   H.~Kowalski, T.~Lappi and R.~Venugopalan,
  {\it ``Nuclear enhancement of universal dynamics of high parton densities,''}
  Phys.\ Rev.\ Lett.\  {\bf 100} (2008) 022303
  [arXiv:0705.3047 [hep-ph]].

\bibitem{SATMOD5}
  H.~Kowalski, T.~Lappi, C.~Marquet and R.~Venugopalan,
  {\it ``Nuclear enhancement and suppression of diffractive structure functions at high energies,''}
  Phys.\ Rev.\   {\bf C78} (2008) 045201
  [arXiv:0805.4071 [hep-ph]].
   
\bibitem{SATMOD6}
   G.~Watt and H.~Kowalski,
  {\it ``Impact parameter dependent colour glass condensate dipole model,''}
  Phys.\ Rev.\   {\bf D78} (2008) 014016
  [arXiv:0712.2670 [hep-ph]].
  \bibitem{SATMOD7}
   E.~Levin and A.~H.~Rezaeian,
  {\it ``Gluon saturation and inclusive hadron production at LHC,''}
  Phys.\ Rev.\   {\bf D82} (2010) 014022
  [arXiv:1005.0631 [hep-ph]].
  \bibitem{SATMOD8}
   A.~H.~Rezaeian,
  {\it ``CGC predictions for p+A collisions at the LHC and signature of QCD saturation,''}
  Phys.\ Lett.\  {\bf B718} (2013) 1058
  [arXiv:1210.2385 [hep-ph]].
  
  \bibitem{SATMOD9}
   E.~Levin and A.~H.~Rezaeian,
  {\it ``Gluon saturation and energy dependence of hadron multiplicity in pp and AA collisions at the LHC,''}
  Phys.\ Rev.\  {\bf D83} (2011) 114001
  [arXiv:1102.2385 [hep-ph]].

   \bibitem{SATMOD10}
  E.~Levin and A.~H.~Rezaeian,
 {\it ``Hadron multiplicity in pp and AA collisions at LHC from the Color Glass Condensate,''}
  Phys.\ Rev.\   {\bf D82} (2010) 054003
  [arXiv:1007.2430 [hep-ph]].
  
  \bibitem{SATMOD11}
   D.~Boer, M.~Diehl, R.~Milner, R.~Venugopalan, W.~Vogelsang, D.~Kaplan, H.~Montgomery and S.~Vigdor {\it et al.},
  {\it ``Gluons and the quark sea at high energies: Distributions, polarization, tomography,''}
  [arXiv:1108.1713 [nucl-th]].

  \bibitem{SATMOD12}
   T.~Lappi and H.~Mantysaari,
  {\it ``Incoherent diffractive J/Psi-production in high energy nuclear DIS,''}
  Phys.\ Rev.\   {\bf C83} (2011) 065202
  [arXiv:1011.1988 [hep-ph]].

   \bibitem{SATMOD13}
   T.~Toll and T.~Ullrich,
  {\it ``Exclusive diffractive processes in electron-ion collisions,''}
  Phys.\ Rev.\   {\bf C87} (2013) 2,  024913
  [arXiv:1211.3048 [hep-ph]].
  
  \bibitem{SATMOD14}
  P.~Tribedy and R.~Venugopalan,
  {\it ``Saturation models of HERA DIS data and inclusive hadron distributions in p+p collisions at the LHC,''}
  Nucl.\ Phys.\  {\bf A850} (2011) 136
   [Nucl.\ Phys.\ A {\bf 859} (2011) 185]
  [arXiv:1011.1895 [hep-ph]].
  
  \bibitem{SATMOD15}
    P.~Tribedy and R.~Venugopalan,
  {\it ``QCD saturation at the LHC: comparisons of models to p+p and A+A data and predictions for p+Pb collisions,''}
  Phys.\ Lett.\   {\bf B710} (2012) 125
   [Phys.\ Lett.\   {\bf B718} (2013) 1154]
  [arXiv:1112.2445 [hep-ph]].
  
 \bibitem{SATMOD16}
  A.~H.~Rezaeian, M.~Siddikov, M.~Van de Klundert and R.~Venugopalan,
  {\it ``IP-Sat: Impact-Parameter dependent Saturation model  revised,''}
  PoS DIS {\bf 2013} (2013) 060
  [arXiv:1307.0165 [hep-ph]];\,\,\, 
  {\it ``Analysis of combined HERA data in the Impact-Parameter dependent Saturation model,''}
  Phys.\ Rev.\   {\bf D87} (2013) 3,  034002
  [arXiv:1212.2974].

  \bibitem{SATMOD17}
 A.~H.~Rezaeian and I.~Schmidt,
  {\it ``Impact-parameter dependent Color Glass Condensate dipole model and new combined HERA data,''}
  Phys.\ Rev.\   {\bf D88} (2013) 074016
  [arXiv:1307.0825 [hep-ph]].

\bibitem{CLP}
  C.~Contreras, E.~Levin and I.~Potashnikova,
{\it ``CGC/saturation approach: a new impact-parameter dependent model,''}
Nucl. Phys.   \textbf{A948} (2016), 1-18
[arXiv:1508.02544 [hep-ph]].

\bibitem{CLS}
C.~Contreras, E.~Levin and M.~Sanhueza,
Phys. Rev. D \textbf{104} (2021) no.11, 116020
[arXiv:2106.06214 [hep-ph]].

\bibitem{LEBR}
G.~P.~Lepage and S.~J.~Brodsky,
Phys. \ Rev.\ Lett.\ {\bf 43} (1979) 545; Phys. \ Rev. \ Lett. \ {\bf 43} 
(1979) 1625.

\bibitem{HERA1}
 F.~D.~Aaron {\it et al.}  [H1 and ZEUS Collaborations],
  {\it ``Combined Measurement and QCD Analysis of the Inclusive  $e  p$  Scattering Cross Sections at HERA,''}
  JHEP {\bf 1001} (2010) 109
  [arXiv:0911.0884 [hep-ex]].
  
 \bibitem{HERA2}
H.~Abramowicz {\it et al.}  [H1 and ZEUS Collaborations],
  {\it ``Combination and QCD Analysis of Charm Production Cross Section Measurements in Deep-Inelastic ep Scattering at HERA,''}
  Eur.\ Phys.\ J.\ C {\bf 73} (2013) 2,  2311
  [arXiv:1211.1182 [hep-ex]].

\bibitem{EIC}
D.~Boer, M.~Diehl, R.~Milner, R.~Venugopalan, W.~Vogelsang, D.~Kaplan, H.~Montgomery, S.~Vigdor, A.~Accardi and E.~C.~Aschenauer, \textit{et al.}
[arXiv:1108.1713 [nucl-th]]; 
R.~Abdul Khalek, U.~D'Alesio, M.~Arratia, A.~Bacchetta, M.~Battaglieri, M.~Begel, M.~Boglione, R.~Boughezal, R.~Boussarie and G.~Bozzi, \textit{et al.}
[arXiv:2203.13199 [hep-ph]].


\bibitem{LHEC}
J.~L.~Abelleira Fernandez \textit{et al.} [LHeC Study Group],
J. Phys. G \textbf{39} (2012), 075001
[arXiv:1206.2913 [physics.acc-ph]].


\bibitem{KUSA}
K.~Kutak and S.~Sapeta,
Phys. Rev. D \textbf{86} (2012), 094043
[arXiv:1205.5035 [hep-ph]].

\bibitem{CLS2}
C.~Contreras, E.~Levin and M.~Sanhueza,
``Soft pomeron in the color glass condensate approach,''
Phys. Rev. D \textbf{106} (2022) no.3, 034011
[arXiv:2203.10296 [hep-ph]].

 \bibitem{BGBP}
J.~Bartels, K.~Golec-Biernat and K.~Peters,
``On the Dipole Picture in the Nonforward Direction,''
Acta Phys. Polon. \textbf{B34}, 3051 (2003) 

\bibitem{HXY}
Y.~Hatta, B.~W.~Xiao and F.~Yuan,
``Gluon Tomography from Deeply Virtual Compton Scattering at Small-x,''
Phys. Rev. D \textbf{95}, no.11, 114026 (2017)
doi:10.1103/PhysRevD.95.114026
[arXiv:1703.02085 [hep-ph]].


\bibitem{GLM}
E.~Gotsman, E.~M.~Levin and U.~Maor,
Z. Phys. C \textbf{57} (1993), 677-684
[arXiv:hep-ph/9209218 [hep-ph]].

\bibitem{NNPZ}
J.~Nemchik, N.~N.~Nikolaev and B.~G.~Zakharov,
Phys. Lett. B \textbf{341} (1994), 228-237
[arXiv:hep-ph/9405355 [hep-ph]];
J.~Nemchik, N.~N.~Nikolaev, E.~Predazzi and B.~G.~Zakharov,
Z. Phys. C \textbf{75} (1997), 71-87
[arXiv:hep-ph/9605231 [hep-ph]]; J.~R.~Forshaw, R.~Sandapen and G.~Shaw,
Phys. Rev. D \textbf{69} (2004), 094013
[arXiv:hep-ph/0312172 [hep-ph]].

\bibitem{MRT}
A.~D.~Martin, M.~G.~Ryskin and T.~Teubner,
Phys. Rev. D \textbf{62} (2000), 014022
[arXiv:hep-ph/9912551 [hep-ph]].

\bibitem{SGBMR}
A.~G.~Shuvaev, K.~J.~Golec-Biernat, A.~D.~Martin and M.~G.~Ryskin,
Phys. Rev. D \textbf{60} (1999), 014015
[arXiv:hep-ph/9902410 [hep-ph]].

\bibitem{LEPP}
  E.~Levin,
{\it ``Dipole-dipole scattering in CGC/saturation approach at high energy: summing Pomeron loops,''}
JHEP \textbf{11} (2013), 039
[arXiv:1308.5052 [hep-ph]].

\bibitem{IIML}
   E.~Iancu, K.~Itakura and L.~McLerran,
  {\it ``Geometric scaling above the saturation scale,''}
  Nucl.\ Phys.\   {\bf A708} (2002) 327
  [hep-ph/0203137].   


\bibitem{MS}
H.~M\"antysaari and B.~Schenke,
Phys. Rev. D \textbf{98} (2018) no.3, 034013
doi:10.1103/PhysRevD.98.034013
[arXiv:1806.06783 [hep-ph]].

\bibitem{S}
B.~Schenke,
Rept. Prog. Phys. \textbf{84} (2021) no.8, 082301
doi:10.1088/1361-6633/ac14c9
[arXiv:2102.11189 [nucl-th]].


\bibitem{KUTOLL}
A.~Kumar and T.~Toll,
Phys. Rev. D \textbf{105} (2022) no.11, 114011
doi:10.1103/PhysRevD.105.114011
[arXiv:2202.06631 [hep-ph]].


\bibitem{SATMOD16A}
H.~M\"antysaari and P.~Zurita,
Phys. Rev. D \textbf{98} (2018), 036002
[arXiv:1804.05311 [hep-ph]].

\bibitem{SATMOD16B}
B.~Sambasivam, T.~Toll and T.~Ullrich,
Phys. Lett. B \textbf{803} (2020), 135277
[arXiv:1910.02899 [hep-ph]].

\bibitem{HERAFL2}
 H.~Abramowicz {\it et al.} [ZEUS Collaboration],
  {\it ``Deep inelastic cross-section measurements at large $y$ with the ZEUS detector at HERA,''}
  Phys.\ Rev.\  {\bf D90} (2014) 7,  072002
  [arXiv:1404.6376 [hep-ex]];\,\,\,
    S.~Chekanov {\it et al.}  [ZEUS Collaboration],
  {\it ``Measurement of the Longitudinal Proton Structure Function at HERA,''}
  Phys.\ Lett.\  {\bf B682} (2009) 8
  [arXiv:0904.1092 [hep-ex]].

\bibitem{H1FL}
V.~Andreev \textit{et al.} [H1],
``Measurement of inclusive $e p$ cross sections at high $Q^2$ at $\sqrt s =$ 225 and 252 GeV and of the longitudinal proton structure function $F_L$ at HERA,''
Eur. Phys. J. C \textbf{74}, no.4, 2814 (2014)
[arXiv:1312.4821 [hep-ex]].


\bibitem{AAMQS}
J.~L.~Albacete, N.~Armesto, J.~G.~Milhano, P.~Quiroga-Arias and C.~A.~Salgado,
Eur. Phys. J. C \textbf{71} (2011), 1705
[arXiv:1012.4408 [hep-ph]].


\bibitem{EXCL1}
S.~Chekanov \textit{et al.} [ZEUS],
PMC Phys. A \textbf{1}, 6 (2007)
[arXiv:0708.1478 [hep-ex]].

\bibitem{EXCL2}
S.~Chekanov \textit{et al.} [ZEUS],
Eur. Phys. J. C \textbf{24}, 345-360 (2002)
[arXiv:hep-ex/0201043 [hep-ex]].

\bibitem{EXCL3}
S.~Chekanov \textit{et al.} [ZEUS],
Nucl. Phys. B \textbf{695}, 3-37 (2004)
[arXiv:hep-ex/0404008 [hep-ex]].

\bibitem{EXCL4}
A.~Aktas \textit{et al.} [H1],
Eur. Phys. J. C \textbf{46}, 585-603 (2006)
[arXiv:hep-ex/0510016 [hep-ex]].

\bibitem{EXCL5}
S.~Chekanov \textit{et al.} [ZEUS],
Nucl. Phys. B \textbf{718}, 3-31 (2005)
[arXiv:hep-ex/0504010 [hep-ex]].

\bibitem{EXCL6}
F.~D.~Aaron \textit{et al.} [H1],
JHEP \textbf{05}, 032 (2010)
[arXiv:0910.5831 [hep-ex]].

\bibitem{EXCL8}
S.~Chekanov \textit{et al.} [ZEUS],
JHEP \textbf{05}, 108 (2009)
[arXiv:0812.2517 [hep-ex]].

\bibitem{EXCL7}
F.~D.~Aaron \textit{et al.} [H1],
Phys. Lett. B \textbf{681}, 391-399 (2009)
[arXiv:0907.5289 [hep-ex]];\,\,\,
S.Chekanov  {\it et al.}  [ZEUS Collaboration],
PMC Phys. {\bf A1} , 6 (2007),[arXiv:0812.2517 [hep-ex]].

\bibitem{FS04}
J.~R.~Forshaw and G.~Shaw,
JHEP \textbf{12} (2004), 052
[arXiv:hep-ph/0411337 [hep-ph]].

\end{thebibliography}
\end{document}